\newcommand{\CR}{\hbox{{$\cal R$}}} 
\newcommand{\CF}{\hbox{{$\cal F$}}}
\newcommand{\CH}{\hbox{{$\cal H$}}}
\newcommand{\CG}{\hbox{{$\cal G$}}}
\newcommand{\CC}{\hbox{{$\cal C$}}}
\newcommand{\CA}{\hbox{{$\cal A$}}}

\newcommand{\cm}{\mathfrak{m}}  
\newcommand{\cg}{\mathfrak{g}}
\newcommand{\cx}{\mathfrak{x}}

\newcommand{\R}{\mathbb{R}}
\newcommand{\C}{\mathbb{C}}
\newcommand{\Z}{\mathbb{Z}}
\newcommand{\quat}{\mathbb{H}}

\newcommand{\h}{{\scriptstyle\frac{1}{2}}}
\newcommand{\extd}{{\rm d}}
\newcommand{\del}{\partial}
\newcommand{\isom}{{\cong}}
\newcommand{\eps}{{\epsilon}}
\newcommand{\tens}{\mathop{\otimes}}
\newcommand{\la}{{\triangleright}}
\newcommand{\ra}{{\triangleleft}}
\newcommand{\grav}{\mathsf{G}}
\newcommand{\Ext}{{\rm Ext}}
\newcommand{\id}{{\rm id}}
\newcommand{\<}{\langle}
\renewcommand{\>}{\rangle}

\newcommand{\link}{{\rm link}}
\newcommand{\from}{{\longleftarrow}}

\newcommand{\und}[1]{{\underline {#1}}}
\newcommand{\eqn}[2]{\begin{equation}#2\label{#1}\end{equation}}

\newcommand{\rcross}{{\triangleright\!\!\!<}}
\newcommand{\lcross}{{>\!\!\!\triangleleft}}
\newcommand{\cobicross}{{\triangleright\!\!\!\blacktriangleleft}}
\newcommand{\bicross}{{\blacktriangleright\!\!\!\triangleleft}}
\newcommand{\dcross}{{\bowtie}}
\newcommand{\lbiprod}{{>\!\!\!\triangleleft\kern-.33em\cdot}}
\newcommand{\rbiprod}{{\cdot\kern-.33em\triangleright\!\!\!<}}

\documentclass[11pt]{article}
\usepackage{amssymb,amsmath,epsfig}
\begin{document}\baselineskip 18pt

{\ }\qquad \hskip 4.3in
\vspace{.2in}

\begin{center} {\LARGE QUANTUM GROUPS AND NONCOMMUTATIVE GEOMETRY}
\\ \baselineskip 13pt{\ }\\
{\ }\\ Shahn Majid \\ {\ }\\ School of Mathematical Sciences,
Queen Mary and Westfield College\\ University of London, Mile End
Rd, London E1 4NS, UK
\end{center}
\begin{center}
November, 1999
\end{center}

\begin{quote}\baselineskip 13pt
\noindent{\bf Abstract} Quantum groups emerged in the latter
quarter of the 20th century as, on the one hand, a deep and
natural generalisation of symmetry groups for certain integrable
systems, and on the other as part of a generalisation of geometry
itself powerful enough to make sense in the quantum domain. Just
as the last century saw the birth of classical geometry, so the
present century sees at its end the birth of this quantum or
noncommutative geometry, both as an elegant mathematical reality
and in the form of the first theoretical predictions for
Planck-scale physics via ongoing astronomical measurements.
Noncommutativity of spacetime, in particular, amounts to a
postulated new force or physical effect called cogravity.
\end{quote}

\baselineskip 18pt
\section{Introduction}

Now that quantum groups and their associated quantum geometry have
been around for more than a decade, it is surely time to take
stock. Where did quantum groups come from, what have they achieved
and where are they going? This article, which is addressed to
non-specialists (but should also be interesting for experts) tries
to answer this on two levels. First of all on the level of quantum
groups themselves as mathematical tools and building blocks for
physical models. And, equally importantly, quantum groups and
their associated noncommutative geometry in terms of their overall
significance for mathematics and theoretical physics, i.e., at a
more conceptual level. Obviously this latter aspect will be very
much my own perspective, which is that of a theoretical physicist
who came to quantum groups a decade ago as a tool to unify quantum
theory and gravity in an algebraic approach to Planck scale
physics. This is in fact only one of the two main origins in
physics of quantum groups; the other being integrable systems,
which I will try to cover as well. Let me also say that
noncommutative geometry has other approaches, notably the one of
A. Connes coming out of operator theory. I will say something
about this too, although, until recently, this has largely been a
somewhat different approach.

We start with the conceptual significance for theoretical physics.
It seems clear to me that future generations looking back on the
20th century will regard the discovery of quantum mechanics in the
1920s, i.e. the idea to replace the coordinates $x,p$ of classical
mechanics by noncommuting operators $\bf x,p$, as one of its
greatest achievements in our understanding of Nature, matched in
its significance only by the unification of space and time as a
theory of gravity. But whereas the latter was well-founded in the
classical geometry of Newton, Gauss, Riemann and Poincar\'e,
quantum theory was something much more radical and mysterious.
Exactly which variables in the classical theory should correspond
to operators? They are local coordinates on phase space but how
does the {\em global geometry} of the classical theory look in
the quantum theory, what does it fully correspond to? The problem
for most of this century was that the required mathematical
structures to which the classical geometry might correspond had
not been invented and such questions could not be answered.

As I hope to convince the reader, quantum groups and their
associated noncommutative geometry have led in the last decades of
the 20th century to the first definitive answers to this kind of
question. There has in fact emerged a more or less systematic
generalisation of geometry every bit as radical as the step from
Euclidean to non-Euclidean, and powerful enough not to break down
in the quantum domain. I do doubt very much that what we know
today will be the final formulation, but it is a definitive step
in a right and necessary direction and a turning point in the
future development of mathematical and theoretical physics. For
example, any attempt to build a theory of quantum gravity with
classical starting point a smooth manifold -- this includes
loop-variable quantum gravity, string theory and quantum
cosmology, is necessarily misguided except as some kind of
effective approximation: smooth manifolds should {\em come out} of
the algebraic structure of the quantum theory and not be a
starting point for the latter. There is no evidence that the real
world is any kind of smooth continuum manifold except as a
macroscopic approximation and every reason to think that it is
fundamentally not. I therefore doubt that any one of the above
could be a `theory everything' until it becomes an entirely
algebraic theory founded in noncommutative geometry of some kind
or other. Of course, this is my personal view.

At any rate, I do not think that the fundamental importance of
noncommutative geometry can be overestimated. First of all, anyone
who does quantum theory is doing noncommutative geometry whether
wanting to admit it or not, namely noncommutative geometry of the
phase space. Less obvious but also true, we will see in Section~II
that if the position space is curved then the momentum space is by
itself intrinsically noncommutative. If one gets this far then it
is also natural that the position space or spacetime by itself
could be noncommutative, which would correspond to a curved or
nonAbelian momentum group. This is one of the bolder predictions
coming out of noncommutative geometry. It has the simple physical
interpretation as what I call {\em cogravity}, i.e. curvature or
`gravity' {\em in momentum space}. As such it is independent of
i.e. dual to curvature or gravity in spacetime and would appear as
a quite different and new physical effect. Theoretically cogravity
can, for example, be detected as energy-dependence of the speed of
light. Moreover, even if cogravity was very weak, of the order of
a Planck-scale effect, it could still in principle be detected by
astronomical measurements at a cosmological level. {\em Therefore,
just in time for the new millennium, we have the possibility of an
entirely new physical effect in Nature coming from fresh and
conceptually sound new mathematics}.

Where quantum groups precisely come into this is as follows. Just
as Lie groups and their associated homogeneous spaces provided
definitive examples of classical differential geometry even before
Riemann formulated their intrinsic structure as a theory of
manifolds, so quantum groups and their associated quantum
homogeneous spaces, quantum planes etc., provide large (i.e.
infinite) classes of examples of proven mathematical and physical
worth and clear geometrical content on which to build and develop
noncommutative differential geometry. They are noncommutative
spaces in the sense that they have generators or `coordinates'
like the noncommuting operators $\bf x,p$ in quantum mechanics but
with a much richer and more geometric algebraic structure than the
Heisenberg or CCR algebra. In particular, I do not believe that
one can build a theory of noncommutative differential geometry
based on only one example such as the Heisenberg algebra or its
variants (however fascinating) such as the much-studied
noncommutative torus. One needs many more `sample points' in the
form of natural and varied examples to obtain a valid general
theory. By contrast, if one does a search of BIDS one finds, see
Figure~1, vast numbers of papers in which the rich structure and
applications of quantum groups are explored and justified in their
own right (data complied from BIDS: published papers since 1981
with title or abstract containing `quantum group*',`Hopf
alg*',`noncommutative geom*', `braided categ*', `braided group*',
`braided Hopf*'.) This is the significance of quantum groups. And
of course something {\em like} them should be needed in a quantum
world where there is no evidence for a classical space such as
the underlying set of a Lie group.
\begin{figure}
\[\epsfbox{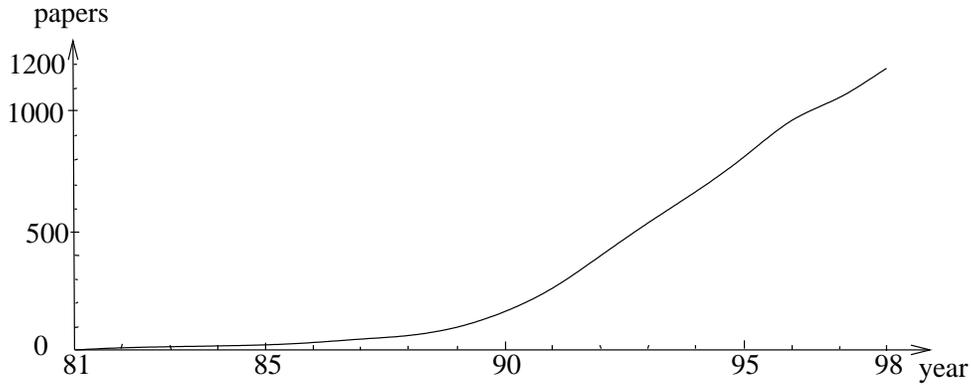}\]
\caption{Growth of research papers on quantum groups}
\end{figure}

Finally, it turns out that noncommutative geometry, at least of
the type that we shall describe, is in many ways cleaner and more
straightforward than the special commutative limit. One simply
does not need to assume commutativity in most geometrical
constructions, including differential calculus and gauge theory.
The noncommutative version is often less infinite, differentials
are often more regular finite-differences, etc. And noncommutative
geometry (unlike classical geometry) can be specialised without
effort to discrete spaces or to finite-dimensional algebras. It is
simply a powerful and natural generalisation of geometry as we
usually know it. So my overall summary and prediction for the next
millennium from this point of view is:
\begin{itemize}
\item All geometry will be noncommutative (or whatever comes
beyond that), with conventional geometry merely a special case.
\item The discovery of quantum theory, its correspondence
principle (and noncommutative geometry is nothing more than the
elaboration of that) will be considered one of the century's
greatest achievement in mathematical physics, commensurate with
the discovery of classical geometry by Newton some centuries
before.
\item Quantum groups will be viewed as the first nontrivial class
of examples and thereby pointers to the correct structure of this
noncommutative geometry.
\item Spacetime too (not only phase space) will be known to be
noncommutative (cogravity will have been detected).
\item At some point a future Einstein will combine the then-standard
noncommutative geometrical ideas with some deep philosophical
ideas and explain something really fundamental about our physical
reality.
\end{itemize}
In the fun spirit of this article, I will not be above putting
down my own thoughts on this last point. These have to do with
what I have called for the last decade the {\em Principle of
representation-theoretic self-duality}\cite{Ma:pri}. In effect, it
amounts to extending the ideas of Born reciprocity, Mach's
principle and Fourier theory to the quantum domain. Roughly
speaking, quantum gravity should be recast as gravity and
cogravity both present and dual to each other and with Einstein's
equation appearing as a self-duality condition. The longer-term
philosophical implications are a Kantian or Hegelian view of the
nature of physical reality, which I propose in Section~V as a new
foundation for next millennium.

We now turn to another fundamental side of quantum groups, which
is at the heart of their other origin in physics, namely as
generalised symmetry groups in exactly solvable lattice models. It
leads to diverse applications ranging from knot theory to
representation theory to Poisson geometry, all areas that quantum
groups have revolutionised. What is really going on here in my
opinion is not so much the noncommutative geometry of quantum
groups themselves as a different kind of noncommutativity or {\em
braid statistics} which certain quantum groups induce on any
objects of which they are a symmetry. The latter is what I have
called `noncommutativity of the second kind' or {\em outer}
noncommutativity since it not so much a noncommutativity of one
algebra as a noncommutative modification of the exchange law or
tensor product of any two independent algebras or systems. It is
the notion of independence which is really being deformed here.
Recall that the other great `isation' idea in mathematical physics
in this century (after `quantisation') was `superisation', where
everything is $\Z_2$-graded and this grading enters into how two
independent systems are interchanged. Physics traditionally has a
division into bosonic or force particles and fermionic or matter
particles according to this grading and exchange behaviour. So
certain quantum groups lead to a generalisation of that as {\em
braided geometry}\cite{Ma:introp} or a process of braidification.
These quantum groups typically have a parameter $q$ and its
meaning is a generalisation of the $-1$ for supersymmetry. This in
turn leads to a profound generalisation of conventional (including
super) mathematics in the form of a new concept of algebra wherin
one `wires up' algebraic operations much as the wiring in a
computer, i.e. outputs of one into inputs of another. Only, this
time, the under or over crossings are nontrivial (and generally
distinct) operations depending on $q$. These are the so-called
`R-matrices'. Afterwards one has the luxury of both viewing $q$ in
this way {\em or} expanding it around $1$ in terms of a multiple
of Planck's constant and calling it a formal `quantisation' --
$q$-deformation actually unifies both `isation' processes. For
example, Lorentz-invariance, by the time it is
$q$-deformed\cite{Ma:varen}, induces braid statistics even when
particles are initially bosonic. In summary,
\begin{itemize}
\item The notion of symmetry or automorphism group is an artifact
of classical geometry and in a quantum world should naturally be
generalised to something more like a quantum group symmetry.
\item Quantum symmetry groups induce braid statistics on the
systems on which they act. In particular,
the notion of bose-fermi statistics or the division into force and
matter particles is an artifact of classical geometry.
\item Quantisation and the departure from bosonic statistics are
two limits of the same phenomenon of braided geometry.
\end{itemize}
Again, there are plenty of concrete models in solid state physics
already known with quantum group symmetry. The symmetry is useful
and can be viewed (albeit with hindsight) as the origin of the
exact solvability of these models.

These two points of view, the noncommutative geometrical and the
generalised symmetry, are to date the two main sources of quantum
groups. One has correspondingly two main flavours or types of
quantum groups which really allowed the theory to take off. Both
were introduced at the mid 1980s although the latter have been
more extensively studied in terms of applications to date. They
include the deformations
\eqn{uqg}{U_q(\cg)}
of the enveloping algebra $U(\cg)$ of every complex semisimple Lie
algebra $\cg$ \cite{Dri}\cite{Jim:dif}. These have as many
generators as the usual ones of the Lie algebra but modified
relations and, additionally, a structure called the `coproduct'.
The general class here is that of {\em quasitriangular quantum
groups}. They arose as generalised symmetries in certain lattice
models but are also visible in the continuum limit quantum field
theories (such as the Wess-Zumino-Novikov-Witten model on the Lie
group $G$ with Lie algebra $\cg$). The coordinate algebras of
these quantum groups are further quantum groups $\C_q[G]$
deforming the commutative algebra of coordinate functions on $G$.
There is again a coproduct, this time expressing the group law or
matrix multiplication. Meanwhile, the type coming out of Planck
scale physics \cite{Ma:pla} are the {\em bicrossproduct quantum
groups}
\eqn{bicgm}{\C[M]\bicross U(\cg)}
associated to the factorisation of a Lie group $X$ into Lie
subgroups, $X=GM$. Here the ingredients are the conventional
enveloping algebra $U(\cg)$ and the commutative coordinate algebra
$\C[M]$. The factorisation is encoded in an action and coaction of
one on the other to make a semidirect product and coproduct
$\bicross$. These quantum arose at about the same time but quite
independently of the $U_q(\cg)$, as the quantum algebras of
observables of certain quantum spaces. Namely it turns out that
$G$ acts on the set $M$ (and vice-versa) and the quantisation of
those orbits are these quantum groups. This means that they are
{\em literally} noncommutative phase spaces of honest quantum
systems. In particular, every complex semisimple $\cg$ has an
associated complexification and its Lie group factorises
$G_{\C}=GG^\star$ (the classical Iwasawa decomposition) so there
is an example
\eqn{bicgg*}{\C[G^\star]\bicross U(\cg)}
built from just the same data as for $U_q(\cg)$. In fact the
Iwasawa decomposition can be understood in Poisson-Lie terms with
$\cg^\star$ the classical `Yang-Baxter dual' of $\cg$. In spite of
this, there is, even after a decade of development, no direct
connection between the two quantum groups:
\[ \cg\]
\eqn{ugbic}{ \swarrow\qquad\searrow}
\[ {\ }\qquad U_q(\cg) \quad \leftarrow \ ?\  \rightarrow
\quad \C[G^\star]\bicross U(\cg).\]
They are both `exponentiations' of the same classical data but
apparently of completely different type (this remains a mystery to
date.)

Associated to these two flavours of quantum groups there are
corresponding homogeneous spaces such as quantum spheres, quantum
spacetimes, etc. Thus, of the first type there is a $q$-Minkowski
space introduced in \cite{CWSSW:ten} as a $q$-Lorentz covariant
algebra, and independently about a year later in \cite{Ma:exa} as
$2\times 2$ braided hermitian matrices. It is characterised by
\eqn{$q$-mink}{[x_i,t]=0,\quad [x_i,x_j]\ne 0.}
Meanwhile, of the second type there is a noncommutative
$\lambda$-Minkowski space with
\eqn{l-mink}{[x_i,t]=\lambda x_i,\quad [x_i,x_j]=0}
which is the one that provides the first known predictions
testable by astronomical measurements (by gamma-ray bursts of
cosmological origin\cite{AmeMa:wav}). This kind of algebra was
proposed as spacetime in \cite{Ma:reg} and in the 4-dimensional
case it was shown in \cite{MaRue:bic} to be covariant under a
Poincar\'e quantum group of bicrossproduct form. These are clearly
in sharp contrast.

\begin{figure}
\[ \epsfbox{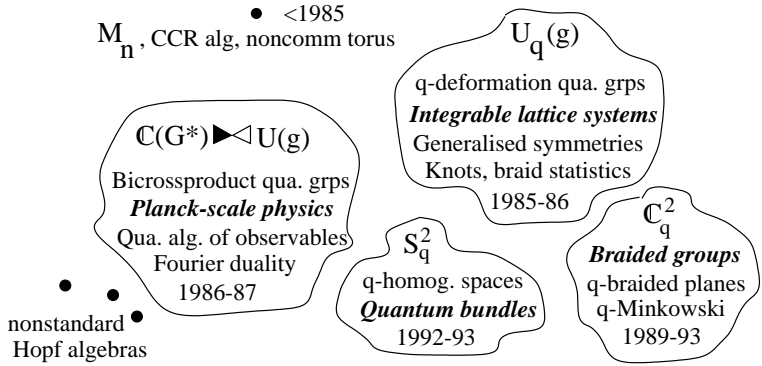}\]
\caption{The landscape of noncommutative geometry today}
\end{figure}

There are of course many more objects than these. $q$-spheres,
$q$-planes etc. In Section~IV we turn to the notion of `quantum
manifold' that is emerging from all these examples. Riemann was
able to formulate the notion of Riemannian manifold as a way to
capture known examples like spheres and tori but broad enough to
formulate general equations for the intrinsic structure of space
itself (or after Einstein, space-time). We are at a similar point
now and what this `quantum groups approach to noncommutative
geometry' is is more or less taking shape. It has the same degree
of `flabbiness' as Riemannian geometry (it is not tied to specific
integrable systems etc.) while at the same time it includes the
`zoo' of already known naturally occurring examples, mostly linked
to quantum groups. Such things as Ricci tensor and Einstein's
equation are not yet understood from this approach, however, so I
would not say it is the last word.

This approach is in fairly sharp contrast to `traditional'
noncommutative geometry as it was done before the emergence of
quantum groups. That theory was developed by mathematicians and
mathematical physicists also coming from quantum mechanics but
being concerned more with topological completions and Hilbert
spaces. Certainly a beautiful theory of von-Neumann and $C^*$
algebras emerged as an analogue of point-set topology. Some
general methods such as cyclic cohomology were also developed in
the 1970s, with remarkable applications throughout mathematics
\cite{Con:geo}. However, for concrete examples with actual
noncommutative {\em differential} geometry one usually turned
either to an actual manifold as input datum or to the Weyl algebra
(or noncommutative torus) defined by relations
\eqn{nctor}{vu=e^{2\pi\imath\theta}uv.} This in turn is basically
the usual CCR or Heisenberg algebra
\eqn{heisxp}{[x,p]=\imath\hbar} in exponentiated form. And at an
algebraic level (i.e. until one considers the precise
$C^*$-algebra completion) this is basically the usual algebra
$B(\CH)$ of operators on a Hilbert space as in quantum mechanics.
Or at roots of unity it is $M_n(\C)$ the algebra of $n\times n$
matrices. So at some level these are all basically one example.
Unfortunately many of the tricks one can pull for this kind of
example are special to it and not a foundation for noncommutative
differential geometry of the type we need. For example, to do
gauge theory Connes and M.~Rieffel\cite{ConRie:yan} used
derivations for two independent vector fields on the torus. The
formulation of `vector field' as a derivation of the coordinate
algebra is what I would call the traditional approach to
noncommutative geometry. For quantum groups such as $\C_q[G]$ one
simply does not have those derivations (rather, they are in
general braided derivations). Similarly, in the traditional
approach one defines a `vector bundle' as a finitely-generated
projective module without any of the {\em infrastructure} of
differential geometry such as a principal bundle to which the
vector bundle might be associated, etc. All of that could not
emerge until quantum groups arrived (one clearly should take a
quantum group as fiber). This is how the quantum groups approach
differs from the work of Connes, Rieffel, Madore and others. It
is also worth noting that string theorists have recently woken up
to the need for a noncommutative spacetime but, so far at least,
have still considered only this `traditional' Heisenberg-type
algebra. In the last year or two there has been some success in
merging these approaches, however; a trend surely to be
continued. By now both approaches have a notion of
`noncommutative manifold' which appear somewhat different but
which have as point of contact the Dirac operator.

\bigskip
\noindent{\em Preliminaries.} A full text on quantum groups is
\cite{Ma:book}. To be self-contained we provide here a quick
definition. Later on we will see many examples and various
justifications for this concept. Thus, a quantum group or {\em
Hopf algebra} is

\begin{itemize}

\item A unital algebra $H,1$ over the field $\C$ (say)

\item A coproduct $\Delta:H\to H\tens H$ and counit $\eps:H\to \C$
forming a {\em coalgebra}, with $\Delta,\eps$ algebra
homomorphisms.

\item An antipode $S:H\to H$ such that
$\cdot(S\tens\id)\Delta=1\eps=\cdot(\id\tens S)\Delta$.
\end{itemize}

Here a coalgebra is just like an algebra but with the axioms
written as maps and arrows on the maps reversed. Thus the
coassociativity and counity axioms are
\eqn{coassoc}{ (\Delta\tens\id)\Delta=(\id\tens\Delta)\Delta,
\quad (\eps\tens\id)\Delta=(\id\tens\eps)\Delta=\id.}
The antipode plays a role that generalises the concept of group
inversion. Other than that the only new mathematical structure
that the reader has to contend with is the coproduct $\Delta$ and
its associated counit. There are several ways of thinking about
the meaning of this depending on our point of view. If the quantum
group is like the enveloping algebra $U(\cg)$ generated by a Lie
algebra $\cg$, one should think of $\Delta$ as providing the rule
by which actions extend to tensor products. Thus, $U(\cg)$ {\em
is} trivially a Hopf algebra with
\eqn{ug}{ \Delta \xi=\xi\tens 1+1\tens \xi,\quad \forall \xi\in \cg,}
which says that when a Lie algebra element $\xi$ acts on tensor
products it does so by $\xi$ in the first factor and then $\xi$ in
the second factor. Similarly it says that when a Lie algebra acts
on an algebra it does so as a derivation. On the other hand, if
the quantum group is like a coordinate algebra $\C[G]$ then
$\Delta$ expresses the group multiplication and $\eps$ the group
identity element $e$. Thus, if $f\in \C[G]$ the coalgebra is
\eqn{C(G)}{ (\Delta f)(g,h)=f(gh),\quad\forall g,h\in G\quad
\eps f=f(e)}
at least for suitable $f$ (or with suitable topological
completions). In other words it expresses the group product
$G\times G\to G$ by a map in the other direction in terms of
coordinate algebras. From yet another point of view $\Delta$
simply makes the dual $H^*$ also into an algebra. So a Hopf
algebra is basically an algebra such that $H^*$ is also an
algebra, in a compatible way, which makes the axioms `self-dual'.
For every finite-dimensional $H$ there is a dual $H^*$. Similarly
in the infinite-dimensional case. It said that in the Roman
empire, `all roads led to Rome'. It is remarkable that several
different ideas for generalising groups all led to the same
axioms. The axioms themselves were first introduced (actually in a
super context) by H. Hopf in 1947 in his study of group cohomology
but the subject only came into its own in the mid 1980s with the
arrival from mathematical physics of the large classes of examples
(as above) that are neither like $U(\cg)$ nor like $\C[G]$, i.e.
going truly beyond Lie theory or algebraic group theory.

\bigskip

\noindent{\em Acknowledgements.} An announcement of this article
appears in a short millennium article\cite{Ma:mil} and a version
more focused on the meaning for Planck scale physics in
\cite{Ma:mea}.

\section{Quantum groups and Planck scale physics}

This section covers quantum groups of the bicrossproduct type
coming out of Planck-scale physics\cite{Ma:pla} and their
associated noncommutative geometry. These are certainly less
well-developed than the more familiar $U_q(\cg)$ in terms of their
concrete applications; one does not have interesting knot
invariants etc. On the other hand, these quantum groups have a
clearer physical meaning as models of Planck scale physics and are
also technically easier to construct. Therefore they are a good
place to start.

Obviously if we want to unify quantum theory and geometry then a
necessary first step is to cast both in the same language, which
for us will be that of algebra. We have already mentioned that
vector fields can be thought of classically as derivations of the
algebra of functions on the manifold, and if one wants points they
can be recovered as maximal ideals in the algebra, etc. This is
the more of less standard idea of algebraic geometry dating from
the late 19th century and early on in the 20th. It will certainly
need to be modified before it works in the noncommutative case but
it is a starting point. The algebraic structure on the quantum
side will need more attention, however.

\subsection{Cogravity}

We begin with some very general considerations. In fact there are
fundamental reasons why one needs noncommutative geometry for any
theory that pretends to be a fundamental one. Since gravity and
quantum theory both work extremely well in their separate domains,
this comment refers mainly to a theory that might hope to unify
the two. As a matter of fact I believe that, through
noncommutative geometry, this `holy grail' of theoretical physics
may now be in sight.

The first point is that we usually do not try to apply or extend
our geometrical intuition to the quantum domain directly, since
the mathematics for that has traditionally not been known. Thus,
one usually considers quantisation as the result of a process
applied to an underlying classical phase space, with all of the
geometrical content there (as a Poisson manifold). But demanding
any algebra such that its commutators to lowest order are some
given Poisson bracket is clearly an illogical and ill-defined
process. It not only does not have a unique answer but also it
depends on the coordinates chosen to map over the quantum
operators. Almost always one takes the Poisson bracket in a
canonical form and the quantisation is the usual CCR or canonical
commutation relations algebra. Maybe this is the local picture but
what of the global geometry of the classical phase space? Clearly
all of these problems are putting the cart before the horse: the
real world is to our best knowledge quantum so that should come
first. We should build models guided by the intrinsic
(noncommutative) geometry at the level of noncommutative algebras
and only at the end consider classical limits and classical
geometry (and Poisson brackets) as emerging from a choice, where
possible, of `classical handles' in the quantum system.

In more physical terms, classical observables should come out of
quantum theory as some kind of limit and not really be the
starting point; in quantum gravity, for example, classical
geometry should appear as an idealisation of the expectation value
of certain operators in certain states of the system. Likewise in
string theory one starts with strings moving in classical
spacetime, defines Lagrangians etc. and tries to quantise. Even in
more algebraic approaches, such as axiomatic quantum field theory,
one still assumes an underlying classical spacetime and classical
Poincar\'e group etc., on which the operator fields live. Yet if
the real world is quantum then phase space and hence probably
spacetime itself should be `fuzzy' and only approximately modeled
by classical geometrical concepts. Why then should one take
classical geometrical concepts inside the functional integral
except other than as an effective theory or approximate model
tailored to the desired classical geometry that we hope to come
out. This can be useful but it cannot possibly be the fundamental
`theory of everything' if it is built in such an illogical manner.
There is simply no evidence for the assumption of nice smooth
manifolds other than now-discredited classical mechanics. And in
certain domains such as, but not only, in Planck scale physics or
quantum gravity, it will certainly be unjustified even as an
approximation.

Next let us observe that any quantum system which contains a
nonAbelian global symmetry group is already crying out for
noncommutative geometry. This is in addition to the more obvious
position-momentum noncommutativity of quantisation. The point is
that if our quantum system has a nonAbelian Lie algebra symmetry,
which is usually the case when the classical system does, then
from among the quantum observables we should be able to realise
the generators of this Lie algebra. That is, the algebra of
observables $A$ should contain the algebra generated by the Lie
algebra,
\eqn{Ag}{A\supseteq U(\cg).}
Typically, $A$ might be the semidirect product of a smaller part
with external symmetry $\cg$ by the action of $U(\cg)$ (which
means that in the bigger algebra the action of $\cg$ is
implemented by the commutator). This may sound fine but if the
algebra $A$ is supposed to be the quantum analogue of the
`functions on phase space', then for part of it we should regard
$U(\cg)$ `up side down' not as an enveloping algebra but as a
noncommutative space with $\cg$ the noncommutative coordinates. In
other words, if we want to elucidate the geometrical content of
the quantum algebra of observables then part of that will be to
understand in what sense $U(\cg)$ is a coordinate algebra,
\eqn{Ug?}{ U(\cg)=\C[?].}
Here $?$ cannot be an ordinary space because its supposed
coordinate algebra $U(\cg)$ is noncommutative.

A concrete example is provided by Mackey quantisation of a
homogeneous space. Thus, if a compact group $G$ acts on a space
$M$ then it induces a metric on it such that the geodesics are
basically the flows under the group action, i.e. particles move on
orbits. The orbits can be quantised in one go as the algebra of
observables given by the cross product
\eqn{mackey}{ \C[M]\lcross U(\cg)}
(in an algebraic setting, say). The natural momentum coordinates
here are the Lie algebra $\cg$ itself and its cross relations with
the position functions $\C[M]$ provide a curved-space analogue of
the Heisenberg commutation relations. There is also a Poisson
structure on $M\times \cg^*$, which is the classical phase space.
$U(\cg)$ is the Kirillov-Kostant quantisation of the $\cg^*$ part.
This class of models is an example of a general principle:
curvature in position space corresponds to noncommutativity of the
natural momentum generators. On a general curved space it means
noncommutativity of covariant derivatives.

So we need noncommutative geometry both for noncommutative phase
space (due to Heisenberg type relations between position and
momentum) and for noncommutative momentum space (when there is
curvature). Finally, since we need a noncommutative geometric
formalism anyway, we may as well allow noncommutative position
space or spacetime too. Only in this way could one restore any
kind of Born reciprocity or symmetry between position and momentum
in the quantum theory. Or more generally, only in this way could
we really imagine canonical transformations mixing position and
momentum coordinates. In other words when Mackey quantisation is
combined with symplectic ideas or with ideas of position-momentum
symmetry one is led naturally to expect that space or spacetime
too should be allowed to be noncommutative.

Let us put these arguments in a more down-to-earth manner. Thus,
in conventional flat space quantum mechanics we take the $\bf x$
commuting among themselves and their momenta $\bf p$ likewise
commuting among themselves. The commutation relation
\eqn{heis}{ [x_i,p_j]=\imath\hbar\delta_{ij}}
is symmetric in the roles of ${\bf x},{\bf p}$ (up to a sign). To
this symmetry may be attributed such things as wave-particle
duality. A wave has localised $\bf p$ and a particle has localised
$\bf x$. Meanwhile, the meaning of curvature in position space is,
roughly speaking, to make the natural conserved $\bf p$
coordinates noncommutative. For example, when the position space
is a 3-sphere the natural momentum is $su_2$. The enveloping
algebra $U(su_2)$ should be there in the quantum algebra of
observables with relations
\eqn{su2p}{ [p_i,p_j]=\frac{\imath}{ R}\eps_{ijk}p_k}
where $R$ is proportional to the radius of curvature of the $S^3$.
By Born-reciprocity then there should be another possibility which
is {\em curvature in momentum space}. It corresponds under Fourier
theory to noncommutativity of position space. For example if the
momentum space were a sphere with $m$ proportional to the radius
of curvature, the natural position space coordinates would
correspondingly have noncommutation relations
\eqn{su2x}{ [x_i,x_j]=\frac{\imath}{ m}\eps_{ijk}x_k.}
Mathematically speaking this is surely a symmetrical and equally
interesting possibility which might have observable consequences.
And if gravity is, loosely-speaking, curvature in position space
or spacetime then this other effect should be called `cogravity'.
In general terms,

\begin{itemize}
\item For systems constrained in position space one has the usual
tools of differential geometry, curvature etc., of the constrained
`surface' in position space or tools for noncommutative algebras
(such as Lie algebras) in momentum space.
\item For systems constrained in momentum space one has
conventional tools of geometry in momentum space or, by Fourier
theory, suitable tools of noncommutative geometry in position
space.
\end{itemize}

Of course, we do not {\em absolutely} need noncommutative geometry
to work effectively with enveloping algebras of Lie algebras. But
if we wish to view (\ref{su2p}),(\ref{su2x}) geometrically as
noncommuting coordinates then we will correspondingly need to
generalise our notion of geometry. What are `vector fields' on
$U(su_2)$? What are differential forms? And so forth. This is what
we have called in \cite{Ma:ista} a `quantum-geometry
transformation' since a quantum symmetry point of view (such as
the angular momentum generators in a quantum system) is turned
`up-side-down' to a geometrical one. These are nontrivial (but
essentially solved) questions. Understanding them, we can proceed
to construct more complex examples of noncommutative geometry
which are neither $U(\cg)$ nor $\C[G]$, i.e. where both quantum
and geometrical effects are unified or where both gravity and
cogravity are present.

Notice also that the three effects exemplified by the three
equations (\ref{heis})--(\ref{su2x}) are all independent. They are
controlled by three different parameters $\hbar,R,m$ (say). Of
course in a full theory of quantum gravity all three effects could
exist together and be unified into a single noncommutative algebra
containing suitable position and momentum modes. Moreover, even if
we do not know the details of the correct theory of quantum
gravity, if we assume that something like Born reciprocity
survives then all three effects indeed {\em should} show up in the
effective theory where we consider almost-particle states with
position and momenta $\bf x,\bf p$. It would require fine tuning
or some special principle to eliminate any one of them.

Finally, our choice of parameter $m$ is suggestive of mass, which
may be appropriate in the case of a mass-shell in momentum space,
but this is not the only source of curvature in momentum space. If
this case is anything to go by, however, it does suggest the
following symmetrical picture
\begin{itemize}
\item Curvature in position spacetime or gravity governs the
background in which test particles are to move. It is related to
its active gravitational mass.

\item Curvature in momentum space or cogravity governs the wave
equation or rules by which a test particle moves even in flat
space. It is related to its passive inertial mass.
\end{itemize}
Although these remarks are all somewhat vague, we see at least
that noncommutative geometrical ideas should in principle help
make precise some of the deepest insights, such as Mach's
principle that motivated Einstein
himself\cite{Ma:phy}\cite{Ma:pri}. This approach to Planck scale
physics based particularly on Fourier theory to extend the
familiar $\bf x,p$ reciprocity to the case of nonAbelian Lie
algebras and beyond was developed by the author in the
1980s\cite{Ma:the}. See also \cite{Ma:ista}\cite{Ma:mea}.

\subsection{Algebraic structure of quantum mechanics}

In the above discussion we have assumed that quantum systems are
described by algebras generated by position and momentum. Here we
will examine this a little more closely. The physical question to
keep in mind is the following: {\em what happens to the geometry
of the classical system when you quantise?}

To see the problem consider what you obtain when you quantise a
sphere or a torus. In usual quantum mechanics one takes the
Hilbert space on position space, e.g. $\CH=L^2(S^2)$ or
$\CH=L^2(T^2)$ and as `algebra of observables' one takes
$A=B(\CH)$ the algebra of all bounded (say) operators. It is
decreed that every self-adjoint such operator $a$ is an observable
of the system and its expectation value in state $|\psi\>\in \CH$
is
\eqn{pure}{ \<a\>_\psi=\<\psi|a|\psi\>.}
The problem with this is that $B(\CH)$ {\em is the same algebra in
all cases}. The quantum system does know about the underlying
geometry of the configuration space or of the phase space in other
ways; the choice of `polarisation' on the phase space or the
choice of Hamiltonian etc. -- such things are generally defined
using the underlying position or phase space geometry -- but the
abstract algebra $B(\CH)$ doesn't know about this. All separable
Hilbert spaces are isomorphic (although not in any natural way) so
their algebras of operators are also all isomorphic. In other
words, whereas in classical mechanics we use extensively the
detailed geometrical structure, such as the choice of phase space
as a symplectic manifold, all of this is not recorded very
directly in the quantum system. One more or less forgets it,
although it resurfaces in relation to the more restricted kinds of
questions (labeled by classical `handles') that one asks in
practice about the quantum system. In other words,
\begin{itemize}
\item The true quantum
algebra of observables should not be the entire algebra $B(\CH)$
but some restricted subalgebra $A\subset B(\CH)$.
\end{itemize}
The choice of this subalgebra is called the {\em kinematic
structure} and it is precisely here that the (noncommutative)
geometry of the classical and quantum system is encoded. This is
somewhat analogous to the idea in geometry that every manifold can
be visualised concretely embedded in some $\R^n$. Not knowing this
and thinking that coordinates $\bf x$ were always globally defined
would miss out on all physical effects that depend on topological
sectors, such as the difference between spheres and tori.

Another way to put this is that by the Darboux theorem all
symplectic manifolds are {\em locally} of the canonical form
$\extd x\wedge \extd p$ for each coordinate pair. Similarly one
should take (\ref{heis}) (which essentially generates all of
$B(\CH)$, one way or another) only locally. The full geometry in
the quantum system is visible only by considering more nontrivial
algebras than this one to bring out the global structure. We
should in fact consider all noncommutative algebras equipped with
certain structures common to all quantum systems, i.e. inspired by
$B(\CH)$ as some kind of local model or canonical example but not
limited to it. The conditions on our algebras should also be
enough to ensure that there {\em is} a Hilbert space around and
that $A$ can be viewed concretely as a subalgebra of operators on
it.

Such a slight generalisation of quantum mechanics which allows
this kinematic structure to be exhibited exists and is quite
well-known in mathematical physics circles. The required algebra
is a {\em von Neumann} algebra or, for a slightly nicer theory, a
$C^*$-algebra. This is an algebra over $\C$ with a $*$ operation
and a norm $||\ ||$ with certain completeness and other
properties. The canonical example is $B(\CH)$ with the operator
norm and $*$ the adjoint operation, and every other is a
subalgebra.

Does this slight generalisation have observable consequences?
Certainly. For example in quantum statistical mechanics one
considers not only state vectors $|\psi\>$ but `density matrices'
or generalised states. These are convex linear combinations of the
projection matrices or expectations associated to state vectors
$|\psi_i\>$ with weights $s_i\ge 0$ and $\sum_i s_i=1$. The
expectation value in such a `mixed state' is
\eqn{dens}{ \<a\>=\sum_i s_i\<\psi_i|a|\psi_i\>}
In general these possibly-mixed states are equivalent to simply
specifying the expectation directly as a linear map $\<\
\>:B(\CH)\to \C$. This map respects the adjoint or $*$ operation on
$B(\CH)$ so that $\<a^*a\>\ge 0$ for all operators $a$ (i.e. a
positive linear functional) and is also continuous with respect to
the operator norm. Such positive linear functionals on $B(\CH)$
are precisely of the above form (\ref{dens}) given by a density
matrix, so this is a complete characterisation of mixed states
with reference only to the algebra $B(\CH)$, its $*$ operation and
its norm. The expectations $\<\ \>_\psi$ associated to ordinary
Hilbert space states are called the `pure states' and are
recovered as the extreme points in the topological space of
positive linear functionals (i.e. those which are not the convex
linear combinations of any others).

Now, if the actual algebra of observables is some subalgebra
$A\subset B(\CH)$ then any positive linear functional on the
latter of course restricts to one on $A$, i.e. defines an
`expectation state' $A\to \C$ which associates numbers, the
expectation values, to each observable $a\in A$. But not
vice-versa, i.e. the algebra $A$ may have perfectly well-defined
expectation states in this sense which are not extendible to all
of $B(\CH)$ in the form (\ref{dens}) of a density matrix.
Conversely, a pure state on $B(\CH)$ given by $|\psi\>\in\CH$
might be mixed when restricted to $A$. The distinction becomes
crucially important for the correct analysis of quantum
thermodynamic systems for example, see \cite{BraRob:ope}.

The analogy with classical geometry is that not every local
construction may be globally defined. If one did not understand
that one would miss such important things as the Bohm-Aharanov
effect, for example. Although I am not an expert on the
`measurement problem' in the philosophy of quantum mechanics it
does not surprise me that one would get into inconsistencies if
one did not realise that the algebra of observables is a
subalgebra of $B(\CH)$. And from our point of view it is precisely
to understand and `picture' the structure of the subalgebra for a
given system that noncommutative geometry steps in. I would also
like to add that the problem of measurement itself is a matter of
matching the quantum system to macroscopic features such as the
position of measuring devices. I would contend that to do this
consistently one first has to know how to identify aspects of
`macroscopic structure' in the quantum system without already
taking the classical limit. Only in this way can one meaningfully
discuss concepts such as partial measurement or the arbitrariness
of the division into measurer and measured. Such an identification
is exactly the task of noncommutative geometry, which deals with
extending our macroscopic intuitions and classical `handles' over
to the quantum system. Put another way, the correspondence
principle in quantum mechanics typically involves choosing local
coordinates like $\bf x,p$ to map over. Its refinement to
correspond more of the global geometry into the quantum world is
the practical task of noncommutative geometry.

The algebraic structure of quantum theory that we have described
here was used by G.W. Mackey and I. Segal in the 1960s and also
became a key ingredient in axiomatic quantum field theory. From
the point of view of noncommutative geometry the turning point was
a theorem of Gelfand and Naimark in the 1940s that every
commutative $C^*$-algebra corresponds to a locally compact
topological space as its algebra of functions vanishing at
infinity. Based on this, one may regarded {\em any} noncommutative
$C^*$-algebra as `noncommutative topological space'. Similarly a
later theorem of Serre and Swann characterised a vector bundle as
a finitely-generated projective module over the algebra of
functions in the commutative case, so one could adopt this in the
noncommutative case too. This led to the operator theory (or
functional analysis) approach to noncommutative geometry developed
further by A. Connes and others, as explained in Section~I. Here
cyclic cohomology reproduces DeRahm cohomology in the commutative
case. More recently Connes has introduced an operator notion of a
`spectral triple' which apparently in the commutative case
reproduces a spin manifold structure and Dirac operator; see
\cite{Con:geo}.

This operator theory approach to noncommutative geometry is
focused in Hilbert spaces, spectral properties of the `Dirac
operator' etc., i.e. comes out of quantum mechanical thinking in a
kind of `top-down' manner. It complements the more algebraic and
`bottom-up' approach coming out of quantum groups which is more
focused in the differential geometry (e.g. $q$-deforming usual
formulae in differential geometry). The two approaches certainly
can and should benefit each other. For example, just because an
elegant construction gives the right answer in the commutative
case does not mean it is the `right' formulation in the
noncommutative case. This can only be known through experience
with concrete examples that one wishes to include in the more
general theory.

\subsection{nonAbelian Fourier theory and cosmological
$\gamma$-ray bursts}

We now begin to use noncommutative geometry and particularly
quantum group technology to make precise some of the ideas in
Section~II.A about position and momentum and their correspondence
through Fourier theory. We need to extend this to the nonAbelian
case.

Fourier theory on $\R$ is of course familiar. Let us recall that
it also works perfectly well for any (locally compact) Abelian
group $G$. The conjugate group $\hat G$ is the set of characters
on $G$. There is a pairing between a character $\chi$ and a group
element $g$, namely to evaluate $\chi(g)$. The Fourier transform
is then
\eqn{FouAb}{ \CF(f)(\chi)=\int_G\extd g f(g)\chi(g) }
and similarly for the inverse with the roles of $G$ and $\hat G$
interchanged. Its key feature is that it turns differential
operators on $G$ into multiplication by functions in $\hat G$.

However, for nonAbelian groups one still has $\hat G$ as the set
of irreducible representations but it does not form a group any
more and it does not carry enough information to reconstruct the
original group i.e. to allow Fourier transform. The latter is
possible but the Fourier transformed functions are not functions
on $\hat G$ exactly but rather they are matrix-valued `functions'
where the value at $\rho\in\hat G$ lies in ${\rm End}( V_\rho)$,
where $V_\rho$ is the vector space of the representation $\rho$
and where the function is constrained to be consistent with all
(iso)morphisms between different $\rho$. Put another way, one has
to work with the entire category of representations {\em and} the
morphisms between them, not only the set $\hat G$. This is not a
bad point of view and we will return to it in Section~III for
quasitriangular quantum groups, but it is not a very geometrical
one.

Quantum groups provide a more geometrical alternative to this
which keeps the flavour of the Abelian case but at the price of
generalising our notions to noncommutative geometry. Thus, for any
Hopf algebra $H$, recall from (\ref{C(G)}) that if we think of it
as like `functions on a group' then the coproduct corresponds to
the group product law. Hence a translation-invariant integral
means in general a map $\int:H\to \C$ such that
\eqn{int}{ (\int\tens\id)\Delta=1\int. }
Meanwhile, the notion of plane wave or exponential should be
replaced by the canonical element
\eqn{exp}{ \exp=\sum_a e_a\tens f^a\in H\tens H^*}
where $\{e_a\}$ is a basis and $\{f^a\}$ is a dual basis. We can
then define Fourier transform as
\eqn{fou}{ \CF:H\to H^*,\quad \CF(h)=\int (\exp) h
=(\int\sum_a e_a h)f^a.}
There is a similar formula for the inverse $H^*\to H$. In the
infinite-dimensional case on will need bases of our two mutually
dual Hopf algebras and either formal powerseries or a topological
completion of the tensor product (i.e. some real analysis) for
this to make sense.

First of all, we check that we recover usual Fourier theory at
least at some formal level. Thus, take $H=\C[x]$ the algebra of
polynomials in one variable, as the coordinate algebra of $\R$. It
forms a Hopf algebra with
\eqn{Rcoprod}{\Delta x=x\tens 1+1\tens x, \quad \eps x=0\,\quad
Sx=-x}
as an expression of the additive group structure on $\R$.
Similarly we take $\C[p]$ for the coordinate algebra of another
copy of $\R$ with generator $p$ dual to $x$ (the additive group
$\R$ is self-dual). The two Hopf algebras $H=\C[x]$ and
$H^*=\C[p]$ are dual to each other with
$\<x^n,p^m\>=(-\imath)^n\delta_{n,m}n!$ (under which the coproduct
of one is dual to the product of the other). The (formal) exp
element and Fourier transform is therefore
\eqn{expline}{ \exp=\sum\imath^n\frac{x^n\tens p^n}{n!}
=e^{\imath x\tens p},\quad \CF(f)(p)=\int_{-\infty}^\infty \extd x
f(x)e^{\imath x\tens p}.} Apart from an implicit $\tens$ symbol
which one does not usually write, we recover usual Fourier theory.
Both the notion of duality and the exponential series are being
treated a bit formally but can be made precise, of course.

On the other hand we can apply the formalism just as well to
$H=\C[G]$ the coordinate algebra of a nonAbelian complex Lie group
(for the real forms one afterwards introduces a $*$-operation on
the algebra). These can typically be understood concretely as
matrix groups with $\C[G]$ generated by the coordinate functions
$t^i{}_j$ which assign to a group element its $ij$ matrix entry,
modulo some relations (and afterwards we can take topological
completions). Their coproduct according to (\ref{C(G)}) is
\eqn{deltamat}{ \Delta t^i{}_j=t^i{}_k\tens t^k{}_j}
corresponding to the matrix multiplication or group law. This
quantum group is dual to the enveloping algebra $U(\cg)$ of the
associated Lie algebra $\cg$ with duality pairing
\eqn{pairingmat}{ \<t^i{}_j,\xi\>=\rho(\xi)^i{}_j,}
where $\rho$ is the corresponding matrix representation of the Lie
algebra. The canonical element or $\exp$ is given by choosing a
basis for $U(\cg)$ and finding its dual basis. We do have
integrals and hence, at least formally, a Fourier transform
\eqn{FouUg}{ \CF:\C[G]\to U(\cg)}
and back. The action of vector fields given by elements of $\cg$
become multiplication in $U(\cg)$, etc. In the reverse direction
we have to take the view that $U(\cg)$ is a noncommutative space
and find an integral on it, etc. But since it is a perfectly good
Hopf algebra we have no problem in doing any of this or in proving
the usual properties of Fourier theory. Thus there are `vector
fields' in $U(\cg)$ given by the action of the $t^i{}_j$ and they
Fourier transform to multiplication in $\C[G]$, etc.

For example, one could apply this to $H=\C[SU_2]=\C[a,b,c,d]$
modulo the relation $ad-bc=1$ (and a $*$-operation to express
unitarity). It has coproduct
\eqn{deltasu2}{ \Delta a=a\tens a+b\tens c,\quad {\rm etc.},\quad
\Delta
\begin{pmatrix}a & b\\ c &d\end{pmatrix}
= \begin{pmatrix}a & b\\ c &d\end{pmatrix} \tens
\begin{pmatrix}a & b\\ c &d\end{pmatrix}.}
The duality pairing with $U(su_2)$ in its usual antihermitian
generators $\{e_i\}$ is \eqn{pairingsu2}{ \<\begin{pmatrix}a &
b\\ c &d\end{pmatrix},e_i\> =\frac{\imath}{2}\sigma_i,} defined
by the Pauli matrices. Let $\{e_1^ae_2^be_3^c\}$ be a basis of
$U(su_2)$ and $\{f^{a,b,c}\}$ the dual basis, then we have a
Fourier transform \eqn{Fousu2}{ \CF:\C[SU_2]\to U(su_2),\quad
\CF(f) =\left(\int_{SU_2}\extd u f(u)f^{a,b,c}(u)\right) e_1^a
e_2^b e_3^c.} Here $\extd u$ denotes the right-invariant Haar
measure on $SU_2$. For a picture of $e_i$ as the coordinates on
momentum space conjugate to position space $SU_2$, we have to
regard $U(su_2)$ as coordinates of a `noncommutative space'. Or
we could equally well reverse the roles of these quantum groups
(i.e. focus on the inverse Fourier transform) as connecting
noncommutative position space with coordinates $U(su_2)$ and
commutative but curved $SU_2$ momentum space. Note that one can
certainly put in the functional analysis in both cases. For
example, as $C^*$-algebras the role of $U(\cg)$ is more properly
played by the group $C^*$-algebra $C^*(G)$, etc.

For an even more concrete example we take the Lie algebra
$\R\lcross\R$ with generators $x,t$ and relations
\eqn{l-mink2}{[x,t]=\imath\lambda
x.} Its enveloping algebra $U(\R\lcross\R)$ could be viewed as a
noncommutative analogue of 1+1 dimensional space-time. Its
associated Lie group $\R\lcross\R$ consists of matrices of the
form
\eqn{Minkmat}{\left(\begin{matrix}e^{\lambda\omega} & k\\ 0
&1\end{matrix}\right)}
and has coordinate algebra $\C[\R\lcross\R]=\C[k,\omega]$ with
coproduct
\eqn{deltaminkmat}{ \Delta e^{\lambda\omega}=e^{\lambda\omega}
\tens e^{\lambda\omega},
\quad\Delta k=k\tens 1+e^{\lambda\omega}\tens k }
Its duality pairing with $U(\R\lcross\R)$ is generated by
$\<x,k\>=-\imath,
\<t,\omega\>=-\imath$ and the resulting exp and Fourier transform
are\cite{MaOec:twi}
\eqn{Foumink}{ \exp=e^{\imath k\omega}e^{\imath\omega t},\quad
\CF(:f(x,t):)
=\int_{-\infty}^\infty\int_{-\infty}^\infty \extd x
\extd t\ e^{\imath k x}e^{\imath\omega t}f(e^{\lambda\omega}x,t)}
where $:f(x,t):\in U(\R\lcross\R)$ by normal ordering $x$ to the
left of $t$. This is the Fourier transform from $U(\R\lcross\R)$
as a noncommutative spacetime to $\C[\R\lcross\R]$ where
$\R\lcross\R$ is the nonAbelian or `curved' momentum space. One
has to complete it in a suitable way of course.

This kind of algebra for spacetime (and its $q$-deformation, but
one may set $q=1$) was proposed by the author in \cite{Ma:reg},
where the relevant integration needed for the Fourier theory was
given. In $3+1$ dimensions the same Lie algebra with $x$ replaced
by a vector $\vec x$ was introduced in \cite{MaRue:bic} as the
suitable spacetime covariantly acted upon by one of the candidates
for a deformed Poincar\'e quantum group, as we will see later.
Using the methods above one can view this noncommutative spacetime
$U(\R^3\lcross\R)$ equivalently under Fourier theory as a theory
of nonAbelian momentum group $\R^3\lcross\R$. This and its
physical consequences were explored recently in \cite{AmeMa:wav}.
In particular, one is then able to justify the dispersion relation
\eqn{disp}{ \lambda^{-2}\left(e^{\lambda\omega}
+e^{-\lambda\omega}-2\right)
-\vec k^2 e^{-\lambda\omega}=m^2}
as a well-defined mass-shell in the classical momentum group
$\R^3\lcross\R$ and give some arguments (for the first time) that
the plane waves being of the form $e^{\imath \vec k\cdot\vec
x}e^{\imath
\omega t}$ above would have wave velocities given by
$v_i=\frac{\del \omega}{\del k_i}$. One then arrives at the
prediction that the velocity of such noncommutative waves (under a
lot of hypotheses concerning how they might be measured) would
depend on energy as
\eqn{pred}{|v|=e^{-\omega\lambda}.}
Some heuristic speculations of this type first appeared in
\cite{ALN:def} from thinking about the Casimir in the deformed
Poincar\'e quantum group (one needs the nonAbelian Fourier theory,
however, to connect this with waves in spacetime and justify
$\omega$ as energy etc.)

This theoretical prediction can actually be measured for gamma ray
bursts that travel cosmological distances, even if $\lambda$ is
very small, of the order of the Planck scale. The known gamma-rays
occur in a spread of energies from 0.1-100 Mev and are known to
travel cosmological distances. Hence the accumulated difference in
their arrival times
\eqn{gammatime}{\delta t=\lambda\frac{L}{c}e^{\lambda\omega}
\delta\omega}
due to the above effect could be of the order of milliseconds,
which is observable. Of course, one does not know how much of the
actual spread in arrival times is due to the effect and how much
is part of the initial structure. For this one needs to know the
distance $ L$ over many bursts and use the predicted
$L$-dependence to filter out other effects and our lack of
knowledge of the initial spectrum. Such better data should,
however, be forthcoming in the near future. It was also
conjectured in \cite{AmeMa:wav} that the nonAbelianness of the
momentum group shows up as CPT violation and might be detected by
ongoing neutral-kaon system experiments. Of course, there is
nothing stopping one doing field theory in the form of Feynman
rules on our classical momentum group either, except that one has
to make sense of the meaning of nonAbelianness in the addition of
momentum. These are the first and I believe at the moment the only
concretely testable predictions for Planck scale physics coming
out of noncommutative geometry, in contrast to theoretical and
conceptual ideas.

\subsection{Fourier theory and loop variables}

Fourier theory also ties up with other approaches to quantum
gravity such as the loop variable one that grew out of the
approach to QCD in the 1970s based on Wilson loops. And it gives
another point of view on the deep connection between Chern-Simons
theory, the Wess-Zumino-Novikov-Witten CFT and the knot invariants
related to quantum groups. We will say more about the latter in
Section~III but for the moment we offer a different and more
heuristic point of view based on Fourier transform. As far as I
now it is due to the author in
\cite{Ma:pho}\cite{Ma:fou}\cite{Ma:csta} and some aspects, such as
the regularised linking number, have certainly turned up in modern
developments in loop variable quantum gravity. We refer to
\cite{Ma:mea} for more details.

We will discuss only the Abelian or $U(1)$ theory in any depth,
leaving as open the problem of really pushing through these ideas
in the nonAbelian case. The required groups are not, however,
locally compact so there is no well-defined translation-invariant
measure. However, this is a matter of regularisation and as
physicists we can also apply these ideas formally by pretending
that there is such a measure. With this caveat, the elements
$\kappa$ of the group are disjoint unions of oriented knots (i.e.
links) with a product law that consists of erasing any overlapping
segments of opposite orientation. The dual group is $\CA/\CG$ of
$U(1)$ bundles and (distributional) connections $A$ on them. Thus
given any bundle and connection, the corresponding character is
the holonomy \eqn{holonomy}{ \chi_A(\kappa)=e^{\imath\int_\kappa
A}.} The inverse Fourier transform of some well-known functions on
$\CA/\CG$ as functions on the group of knots are, \eqn{foucs}{
\CF^{-1}({\rm CS})(\kappa)=\int \extd A \ {\rm
CS(A)}e^{-\imath\int_\kappa A} =e^{\frac{\imath}{2\alpha}{\rm
link}(\kappa,\kappa)}} \eqn{foumax}{ \CF^{-1}({\rm Max})(\kappa)
=\int \extd A \ {\rm Max(A)}e^{-\imath\int_\kappa A}
=e^{\frac{\imath}{2\beta}{\rm ind}(\kappa,\kappa)}} where
\eqn{CSMAX}{ {\rm CS}(A)=e^{\frac{\alpha\imath}{2}\int A\wedge
\extd A},\quad {\rm Max}(A) =e^{\frac{\beta\imath}{2}\int
{}^*\extd A\wedge\extd A}} are the Chern-Simmons and Maxwell
actions, $\rm link$ denotes linking number, $\rm ind$ denotes
mutual inductance. The diagonal $\rm ind(\kappa,\kappa)$ is the
{\em mutual self-inductance} i.e. you can literally cut the knot,
put a capacitor and measure the resonant frequency to measure it.
By the way, to make sense of this one has to use a wire of a
finite thickness -- the self-inductance has a log divergence.
This is also the log-divergence of Maxwell theory when one tries
to make sense of the functional integral, i.e. renormalisation
has a clear physical meaning in this context. Meanwhile, the
diagonal $\rm link(\kappa,\kappa)$ is the self-linking number of
a knot with itself, where $\link(\kappa,\kappa')$ between two
possibly intersecting knots is defined as the limit $\eps\to 0$
of the {\em regularised linking number}\cite{Ma:fou}
\eqn{reglink}{ {\rm link}_\eps(\kappa,\kappa')
=\int_{||\vec\eps||<\eps}\extd^3\vec\eps\ \link
(\kappa,\kappa'_{\vec \eps}).} Here $\kappa'_{\vec \eps}$ is the
second knot displaced by the vector $\vec\eps$. This
$\link(\kappa,\kappa')$ gives, for example, $\pm 1/2$ for each
transversal intersection. One can also define it by the Gauss
formula, which is part of the proof of the above results. Note
also that this point of view is {\em distributional} because, as
well as considering honest smooth connections, one considers
`connections' defined entirely by their holonomy. In particular,
given a knot $\kappa$ one may define the distribution $A_\kappa$
by its character as \eqn{dualconn}{
e^{\imath\int_{\kappa'}A_\kappa} =e^{\imath{\rm
link}(\kappa,\kappa')}.} Such distributions are quite
interesting. For example, if one formally evaluates the Maxwell
action on these one has\cite{Ma:pho}\cite{Ma:csta} \eqn{string}{
{\rm Max}(A_\kappa)=e^{\frac{\imath}{4\beta}
\delta^2(0)\int_\kappa \extd t \dot\kappa\cdot\dot\kappa},} the
Polyakov string action. In other words, string theory can be
embedded into Maxwell theory by constraining the functional
integral to such `vortex' configurations. An additional
Chern-Simons term becomes similarly a `topological mass term'
$\link(\kappa,\kappa)$ that could be added to the Polyakov action.

Now the point is that on the side of $\CA/\CG$ there is no problem
passing to the nonAbelian case and no problem writing down
Yang-Mills and Chern-Simons functionals on this space. What are
their Fourier transforms? In the $SU_2$ case the inverse Fourier
transform of the Chern-Simons functional should surely be some
kind of Jones invariant in place of self-linking number. Actually
the Jones invariant is connected only to the fundamental
representation of $SU_2$ but in a conventional setting (see later
sections) there is such a knot invariant for every representation.
Now, just as gauge fields are something like $U(1)$ fields
`tensored' by $U(\cg)$, the dual should be loops `tensored' by the
dual of $U(\cg)$. In conventional terms this would be therefore
loops labelled by representations in $\hat G$ and the universal
Jones invariant would indeed be a functional on this. There are
many complications here that we have glossed over, i.e. this is a
somewhat heuristic picture. First of all, as written above the
group of loops is not sensitive to under or over crossings -- this
enters in the regularisation needed to make sense of the theory as
well as in the definition of self-linking number in the answer.
The same has to be done in the nonAbelian case. Secondly, there is
not a simple `tensor product' here but rather the Lie algebra and
the differential structure of the gauge field are nontrivially
mixed up (a gauge field is not simply a Lie-algebra valued
1-form). At least at the heuristic level, however, this is one way
of thinking about knot invariants\cite{Ma:fou}. Presumably the
same approaches can be applied to Fourier transform the Yang-Mills
action to understand QCD and presumably also to understand
loop-variable quantum gravity. Finally, our results above suggest
a noncommutative geometrical formulation in which the loops would
have values not in $\hat G$ but in $U(\cg)$ regarded as a
noncommutative space. In other words,
\begin{itemize}
\item NonAbelian gauge theories
should be more or less equivalent under nonAbelian Fourier
transform to a theory of loops with values in a manifold crossed
by a noncommutative space $U(\cg)$.
\end{itemize}
At the time of writing such models and such ideas have yet to be
explored, i.e. this is a conjecture for the future.

There is also another connection with noncommutative geometry.
Thus the CCR's for the gauge field can be equivalently formulated
as
\eqn{canpho}{ [\int_\kappa A,\int_{\Sigma}E]=4\pi\imath
\alpha{\rm link}(\kappa,\del\Sigma)}
which is a signed sum of the points of intersection of the loop
$\kappa$ with the surface $\Sigma$. This is the point of view by
which loop variables were introduced (as an approach to QCD on
lattices) by Mandelstam and others. Now, just as the
noncommutative torus (\ref{nctor}) takes the Heisenberg algebra in
an exponentiated form with relations
\eqn{nctornm}{ v^nu^m=e^{2\pi\imath\theta mn}u^mv^n,}
one has for gauge fields in canonical
quantisation\cite{Ma:pho}\cite{Ma:csta}
\eqn{nonpho}{
v_\kappa u_{\kappa'}=e^{4\pi\imath\alpha{\rm
link}(\kappa,\kappa')}u_{\kappa'}v_\kappa}
 where integers are
replaced by knots or links. Here the physical picture is
\eqn{uvAB}{ u_\kappa=e^{\imath\int_\kappa A},\quad
v_\kappa=e^{\imath\int_\kappa \tilde A}} where $\tilde A$ is a
dual connection such that $E=\extd \tilde A$. So constructing the
$u,v$ is equivalent to constructing some distributional operators
$A,E$ with the usual CCR's. The point here is that CCR's and the
noncommutative torus are but the most elementary examples of
noncommutative geometry (at least at an algebraic level). As we
will see in the next section, there are variations of the CCR
algebra that preserve more of the geometric structure of phase
space in the quantum case; one could envisage similar variants for
quantisation of photons and Yang-Mills fields. This is again
something for the future.

\subsection{The Planck-scale quantum group}

We are now ready to move from simple examples like $U(\cg)$
regarded `up side down' as a noncommutative space to
noncommutative spaces that are genuinely different from both
$U(\cg)$ and its dual $\C[G]$. In fact our goal is to unify these
two. We do this in the category of quantum groups because quantum
groups should be the simplest examples of noncommutative geometry.
In this category we want to have something that really is a
quantum algebra of observables of an honest quantum system and at
the same time preserves something of the geometrical structure of
phase space in the quantum case. In the simplest 1+1-dimensional
model the phase space is $\R^2$ with its additive group structure.
This in turn leads to the usual vector fields etc., on $\R^2$. We
want to be able to keep all that geometry even in the quantum
setting. This line of thinking led in the mid 1980s to the
Planck-scale quantum group $\C[x]\bicross_{\hbar,\grav}\C[p]$
generated by $x,p$ with relations and coproduct\cite{Ma:pla}
\eqn{plankqg}{ [x,p]=\imath\hbar(1-e^{-\frac{x}{\grav}}),\quad
\Delta x=x\tens 1+1\tens x,\quad \Delta p=p\tens
e^{-\frac{x}{\grav}}+1\tens p.}
This should be viewed as some kind of `toy model' or effective
theory of Planck-scale physics with stripped-down degrees of
freedom but incorporating the more fundamental ideas in previous
sections. This is how bicrossproduct quantum groups first
appeared, at least in a modern context.

Notice first of all the quantum flat space $\grav\to 0$ limit. In
any situation where $x$ can effectively be treated as having
values $>0$, i.e. for a certain class of quantum states where the
particle is confined to this region, we clearly have flat space
quantum mechanics with the Heisenberg algebra $[x,p]=\imath\hbar$
as $\grav\to 0$. So this quantum group leads to a modification of
usual quantum mechanics by this parameter $\grav$.

To get some idea of the meaning of this deformation by $\grav$,
suppose that $p$ is the natural conserved momentum and the
Hamiltonian is $h=p^2/2m$ (say). The different $[x,p]$ commutation
relations then correspond to different dynamics. This is the
natural point of view as a dynamical system, but if one prefers an
even more conventional point of view one is free to define $\tilde
p=p(1-e^{-\frac{x}{\grav}})^{-1}$, which then has the usual
canonical commutation relations, with $h$ a certain Hamiltonian
  consisting of $\tilde p^2/2m$ plus derivative interaction
  terms. Either way, one finds
\eqn{plankdyn}{ \dot p=0,\quad \dot
x=\frac{p}{m}\left(1-e^{-\frac{x}{\grav}}\right)+O(\hbar)
=v_\infty\left(1-\frac{1}{1+\frac{x}{\grav}+\cdots}\right)+O(\hbar)}
where we identify $p/m$ to $O(\hbar)$ as the velocity $v_\infty<0$
at $x=\infty$. We see that as the particle approaches the origin
it goes more and more slowly and in fact takes an infinite amount
of time to reach the origin. Compare with the formula in standard
radial in-falling coordinates
\eqn{blackhole}{ \dot x=v_\infty\left(1-\frac{1}
{1+\frac{1}{2}\frac{x}{\grav}}\right)}
for the distance from the event horizon of a Schwarzschild black
hole with
\eqn{Grav}{ \grav= \frac{G_{\rm Newton}M}{c^2},}
where $M$ is the background gravitational mass and $c$ is the
speed of light. Thus the heuristic meaning of $\grav$ in our model
is that it measures the background mass or radius of curvature of
the classical geometry of which our Planck scale Hopf algebra is a
quantisation.

Next we notice the classical limit where $\hbar\to 0$. In this
case we have the commutative algebra of $x,p$ i.e. a classical
space but with coproduct that of the group of matrices of the form
\eqn{plankmat}{ \left(\begin{matrix}e^{-\frac{x}{\grav}} & 0\\ p
&1\end{matrix}\right),}
which is therefore the classical phase space for general $\grav$
of the system. This is the group $\R\rcross\R$ of lower-triangular
matrices equipped with a certain Poisson bracket obtained by
classicalising the above relations. Working a little harder, one
finds that the quantum mechanical limit is valid (the effects of
$\grav$ do not show up within one Compton wavelength) if
\eqn{plankesta}{ mM<<m^2_{\rm Planck},}
while the curved classical limit is valid if
\eqn{plankestb}{ mM>>m^2_{\rm Planck}.}
In the general case both points of view coexist in a unified
structure.

We envisage that this model could appear as some effective limit
of an unknown theory of quantum gravity which to lowest order
would appear as spacetime and conventional mechanics on it. But
actually we can make a much stronger statement, for there is a
sense in which the Planck scale quantum group is not merely {\em
a} quantisation of a certain Poisson space but rather comes from
the intrinsic structure of noncommutative algebras themselves. The
idea is that even if the theory of quantum gravity is unknown, we
can use the intrinsic structure of noncommutative algebras to
classify {\em a priori} different possibilities. This is much as a
phenomenologist might use knowledge of topology or cohomology to
classify different {\em a priori} possible effective Lagrangians
without knowing the full high energy theory.

Specifically, if $H_1$, $H_2$ are two quantum groups there is a
theory of the space ${\rm Ext}_0(H_1,H_2)$ of possible extensions
\eqn{extn}{ 0\to H_1\to E\to H_2\to 0}
by some Hopf algebra $E$ obeying certain conditions. We do not
need to go into the mathematical details here but in general one
can show that $E\isom H_1\bicross H_2$ by a `bicrossproduct' Hopf
algebra construction. Suffice it to say that the conditions are
`self-dual' i.e. the dual of the above extension gives
\eqn{dualextn}{ 0\to H_2^*\to E^*\to H_1^*\to 0}
as another extension dual to the first, in keeping with a
philosophy of self-duality of the category in which we work. We
also note that by $\Ext_0$ we mean quite strong extensions. There
is also a weaker notion that admits the possibilities of cocycles
as well, which we are excluding, i.e. this is only the trivial
sector in a certain nonAbelian cohomology. Then it was found
that\cite{Ma:pla}\cite{Ma:phy}
\eqn{extxp}{{\rm
Ext}_0(\C[x],\C[p])=\R\hbar\oplus
\R\grav,}
a two-parameter space, and that any extension
\eqn{extnxp}{ 0\to \C[x]\to E\to\C[p]\to 0}
of position $\C[x]$ by momentum $\C[p]$ forming a Hopf algebra is
of the bicrossproduct form
$E\isom\C[x]\bicross_{\hbar,\grav}\C[p]$. In physical terms what
we are saying is that if we are given $\C[x]$ the position
coordinate algebra and $\C[p]$ defined {\em a priori} as the
natural momentum coordinate algebra then {\em all possible}
quantum phase spaces built from $x,p$ in a controlled way that
preserves duality ideas (Born reciprocity) and retains the group
structure of classical phase space as a quantum group are of this
form labeled by two parameters $\hbar,\grav$. We have not put
these parameters in by hand -- they are simply the mathematical
possibilities being thrown at us. Also, although one cannot draw
too many conclusions from the analogy with a Schwarzschild
black-hole (given that the toy model here is not even
relativistic), the emergence of a coordinate singularity is again
something that we have not put in by hand. In summary,
\begin{itemize}
\item The model shows how one can be forced to discover both
quantum and gravitational effects from the intrinsic structure of
the theory of noncommutative algebras.
\item Such methods lead to tight constraints on the dynamics, with
features such as coordinate singularities.
\end{itemize}
In particular, it is not possible to make a Hopf algebra for $x,p$
with the correct classical limit in this context without a
coordinate singularity. In fact, solving (\ref{extnxp}) is a
second order differential equation for the possible action of the
momentum generators, playing the role in our toy model of
something like Einstein's equation for the metric.

We can also take a `deep quantum gravity' limit $\hbar,\grav\to
\infty$ in the above, with $\frac{\grav}{\hbar}=\lambda$ held
constant. In this case we obtain \eqn{plankdeepqg}{
[x,p]=\imath\lambda x,\quad \Delta x=x\tens 1+1\tens
x,\quad\Delta p=p\tens 1+1\tens p,} which is the noncommutative
spacetime $U(\R\lcross\R)$ in Section~II.C, where the enveloping
algebra is regarded `up side down' as noncommutative coordinates.
This puts some flesh on the idea that noncommutative spacetime or
cogravity might indeed come out of quantum gravity as some kind
of effective description independent of its details. The time
operator would be built from the momentum modes of that deeper
theory. This connects our quantum group ideas with some of the
first testable predictions as explained in Section~II.C.

There are some other remarkable features of the Planck-scale
quantum group which could give us some qualitative insight into
even more novel phenomena at the Planck scale. The most important
is that it is not only of self-dual type in the sense that its
dual is also a Hopf algebra extension, but it is actually
isomorphic to its own dual,
\eqn{planckdual}{ (\C[x]\bicross_{\hbar,\grav}\C[p])^*\isom
\C[\bar p]\cobicross_{\frac{1}{\hbar},\frac{\grav}{\hbar}}
\C[\bar x],}
where $\C[p]^*=\C[\bar x]$ and $\C[x]^*=\C[\bar p]$ in the sense
of an algebraic pairing. Here $\<p,\bar x\>
=\imath$ etc., which then requires a change of the parameters as
shown to make the identification precise. So there is a dual
theory which has just the same form but the roles of $x,p$
interchanged and different parameter values. This means that
whereas we would look for observables $a\in
\C[x]\bicross\C[p]$ as the algebra of observables and states
$\phi\in\C[\bar p]\cobicross\C[\bar x]$ as the dual linear space,
with $\phi(a)$ the expectation of $a$ in state $\phi$ (See
section~II.B), there is a dual interpretation whereby
\eqn{obsstatdua}{{\rm Expectation}=\phi(a)=a(\phi)}
for the expectation of $\phi$ in `state' $a$ with $\C[\bar
p]\cobicross\C[\bar x]$ the algebra of observables in the dual
theory. More precisely, only self-adjoint elements of the algebra
are observables and only positive functionals are states, and a
state $\phi$ will not be exactly self-adjoint in the dual theory
etc. But the physical self-adjoint elements in the dual theory
will be given by combinations of such states, and vice versa. This
is what I have called {\em observable-state duality}. It was
introduced in \cite{Ma:pla} in the 1980s.

Also conjectured at the time of \cite{Ma:pla} was that this
duality should be related to $T$-duality in string theory. As
evidence is the inversion of the constant $\hbar$. In general
terms coupling inversions are indicative of such dualities. Notice
also that Fourier transform implements this T-duality-like
transformation as
\eqn{Tplank}{ \CF: \C[x]\bicross_{\hbar,\grav}\C[p]\to \C[\bar
p]\cobicross_{\frac{1}{\hbar},\frac{\grav}{\hbar}}\C[\bar x]}
Explicitly, it comes out as\cite{MaOec:twi}
\eqn{planckfou}{ \CF(:f(x,p):)=\int_{-\infty}^\infty
\int_{-\infty}^\infty \extd x\extd p
\ e^{-\imath(\bar p+\frac{\imath}{\grav})x}e^{-\imath\bar x(p+p\la)}
f(x,p),} where
\eqn{pact}{ p\la f=-\imath\hbar(1-e^{-\frac{x}{\grav}})
\frac{\del}{\del x}f}
and $f(x,p)$ is a classical function considered as defining an
element of the Planck-scale quantum group by normal ordering $x$
to the left.

The observable-state duality here is not exactly T-duality in
string theory but has some features like it. On the other hand, it
is done here at the quantum level and not in terms only of
Lagrangians. Rather, this approach suggests,
\begin{itemize}
\item A fundamental theory of physics including quantum
gravity should be defined by some kind of algebraic structure
possessing one or more dualities in a representation-theoretic or
observable-state sense.
\item Classical geometry, Lagrangians, etc. would only appear in
classical limits of the algebraic structure and be related to each
other under the dualit(ies).
\end{itemize}
In the above self-dual model the classical picture in the dual is
the same as in the original model with a change of parameters, but
for more general bicrossproducts the model and its dual model can
be quite different in their classical mechanics. This algebraic
duality point of view was introduced by the author in the 1980s in
\cite{Ma:pri}\cite{Ma:the} but we note that some similar ideas are
beginning to be bandied about by string theorists a decade later
under the name of `M-theory'. This is an unknown theory but deemed
but have different classical limits connected by dualities.

Finally, the Planck-scale quantum group allows us to take a first
look at how classical geometry emerges or conversely how it
corresponds in the quantum theory. For example, an infinitesimal
coproduct defines partial differentials,
\eqn{planckdif}{ \del_p:f(x,p):=\frac{\grav}{\imath\hbar}:(f(x,p)
-f(x,p-\imath\frac{\hbar}{\grav})):,\quad \del_x:f(x,p):
=:\frac{\del}{\del x} f:-\frac{p}{\grav}\del_p:f:}
which shows the effects of $\hbar$ in modifying the geometry.
Differentiation in the $p$ direction becomes `lattice regularised'
albeit a little strangely with an imaginary displacement. In the
$\lambda$-deformed Minkowski space setting where $p=t$ it means
that the Euclidean version of the theory related to the Minkowski
one by a Wick-rotation is being lattice-regularised by the effects
of $\hbar$.

Also note that for fixed $\hbar$ the geometrical picture blows up
when $\grav\to 0$. I.e the usual flat space quantum mechanics CCR
algebra does {\em not} admit a deformation of conventional
differential calculus on $\R^2$. Similarly the associated
differential forms have relations with `functions' $f$ in the
Planck-scale quantum group given by
\eqn{planckfdrel}{ f\extd x=(\extd x)f,\quad f\extd p=(\extd
p)f+\frac{\imath\hbar}{\grav}\extd f,} which blow up as $G\to 0$.
Thus,
\begin{itemize}
\item One needs a small amount of `gravity'
to be present for a geometrical picture in the quantum theory.
\end{itemize}

The higher exterior algebra looks more innocent with
\eqn{planckext}{
\extd x\wedge
\extd x=0,\quad \extd x\wedge \extd p=-\extd p\wedge\extd x,\quad
\extd p\wedge \extd p=0}
Starting with the differential forms and derivatives, one can
proceed to gauge theory, Riemannian structures etc., in some
generality. The above formulae are all from a recent work
\cite{MaOec:twi}, where one also finds `quantum' Poisson brackets
and Hamiltonians (and in principle, Lagrangians) in the full
noncommutative theory. Such tools should help to bridge the gap
between model building via classical Lagrangians, which I
personally do not think can succeed at the Planck scale, and some
of the more noncommutative-algebraic ideas above.

\subsection{General construction of bicrossproducts}

The general construction for bicrossproduct quantum groups is as
follows. Suppose that
\eqn{Xfact}{ X=GM}
is a {\em factorisation of Lie groups}. Then one can show that $G$
acts on the set of $M$ and $M$ acts back on the set of $G$ such
that $X$ is recovered as a double cross product $X\isom G\dcross
M$ (simultaneously by the two acting on each other). The actions
are defined by considering $su$ where $s\in M$ and $u\in G$ are in
the wrong order. By unique factorisation there are elements $s\la
u\in G$ and $s\ra u\in M$ such that $su=(s\la u)(s\ra u)$. In
effect, an action of $G$ on $M$ and a `backreaction' of $M$ on $G$
are created at the same time. This was motivated in \cite{Ma:phy}
from Mach's principle or the general idea that every action has
`equal and opposite' reaction.

Now, quantising the orbits of $G$ in $M$ by this action can be
achieved by the cross product algebra $\C[M]\lcross U(\cg)$ (or
its $C^*$-algebra version). The backreaction of $M$ on $G$ can
similarly be used to made a semidirect coproduct structure and
render us a Hopf algebra
\eqn{facbic}{\C[M]\bicross U(\cg)}
The roles of the two Lie groups is
 symmetric and the dual is
 \eqn{facdual}{(\C[M]\bicross U(\cg))^*=U(\cm)\cobicross \C[G],}
where $\cm$ is the Lie algebra of $M$. What this means that there
are certain
 families of homogeneous spaces (the orbits of one group under
 the other) which come in pairs, with the algebra of observables
 of the quantisation of one being the algebra of expectation
 states of the quantisation of the other.

 This is the general
 construction alluded to in Section~I and we see that it comes
 out of Mach's principle in a quantum mechanical setting. There
 are also plenty of examples. In particular, every
 complexification of a semisimple Lie group factorises into its
compact real form $G$ and a certain solvable group $G^\star$, i.e.
$G_\C=GG^\star$, so there is at least one quantum group
$\C[G^\star]\bicross U(\cg)$ for every Lie algebra. The dual is
another quantum group $U(\cg^\star)\cobicross\C[G]$. Let us note
that some years later, in the mid 1990s, there appeared a similar
proposal for non-Abelian T-duality between a suitable sigma-model
on $G$ and another on $G^\star$ \cite{KliSev:dua} which has a
similar flavour to the above except that it is at the classical
level (not at the level of quantum theory as above) and applies to
the evolution of strings in $G,G^\star$ rather than to points. The
extension of sigma-model T-duality to the full quantum theory (as
well as to general factorisations $GM$) using such methods as
above is a direction for ongoing research at the moment. We refer
to \cite{BegMa:poi} for a first step.

Moreover, general Hopf algebra theory shows that $\C[M]\bicross
U(\cg)$ always acts covariantly on $U(\cm)$, so the latter is
always a noncommutative analogue of the space of fields on which
to build some kind of `Schroedinger' representation for the cross
product. There is also a more conventional Hilbert space
representation. From this point of view one could for example
regard $U(\cm)$ as space or spacetime and $\C[M]\bicross U(\cg)$
as a deformation of the enveloping algebra of a group of motions
on it.

For example, the bicrossproduct quantum group
\eqn{3planck}{ \C[\R^2\lcross\R]\bicross U(su_2)}
was constructed in \cite{Ma:mat}\cite{Ma:hop} (actually as a
Hopf-von Neumann algebra) and acts covariantly on
$U(\R^2\lcross\R)$, which is the 3-dimensional version of the
noncommutative spacetime in Section~II.C. Here
$M=SU_2^\star=\R^2\lcross\R$ consists of 3-vectors $\vec s$ with
third component $s_3>-1$ and with the `curved $\R^3$' nonAbelian
group law
\eqn{R2crossR}{\vec{s}\cdot\vec{t}=\vec{s}+(s_3+1)\vec{t}.}
Its Lie algebra is spanned by $t,x_i$ with relations $[x_i,t]=x_i$
for $i=1,2$ as discussed before. On the group $\R^2\lcross\R$
there is an action of $G=SU_2$ by a deformed rotation. The orbits
are still spheres but non-concentrically nested and accumulating
at $s_3=-1$. This is a dynamical system and (\ref{3planck}) is its
Mackey quantisation as a cross product. We have similar features
as for the Planck-scale quantum group, including some kind of
coordinate singularity as $s_3=-1$. At the same time there is a
`backreaction' of $\R^2\lcross\R$ back on $SU_2$. Therefore the
dual system, related by Fourier theory or observable-state
duality, is of the same form, namely
\eqn{dual3planck}{ U(\R^2\lcross\R)\cobicross \C[SU_2].}
It consists of a particle on $SU_2$ moving under the action of
$\R^2\lcross\R$. Moreover, it has a representation as motions in
$U(su_2)$ as some kind of noncommutative space if one wants to
take that point of view.

These constructions and examples were introduced by the author in
the 1980s. Since then several other examples have proven of
specific interest. First of all, the $3+1$-dimensional version
$U(\R^3\lcross\R)$ of the above can be viewed as noncommutative
Minkowski space covariant under the bicrossproduct quantum group
$\C[\R^3\lcross\R]\bicross U(so_{3,1})$ \cite{MaRue:bic}. It was
given explicitly in this work as the usual translation and Lorentz
rotation and boost $p_\mu,M_i,N_i$ and the cross relations and
coproduct
\eqn{lppinca}{ [p_0,M_i]=0,\quad
[p_i,M_j]=\eps_{ijk}p_k,\quad [p_0,N_i]=-p_i,}
\eqn{lplankb}{ [p_i,N_j]=-\delta_{ij}(\frac{1-e^{-{2\lambda p_0}}}{
2\lambda}+\frac{\lambda}{2}{\vec p}^2)+\lambda p_ip_j,}
\eqn{lplankc}{ \Delta N_i=N_i\tens 1+e^{-{\lambda p_0}}\tens
N_i+{\lambda\eps_{ijk}}p_j\tens M_k,\quad \Delta p_i=p_i\tens
1+e^{-{\lambda p_0}}\tens p_i} and $p_0,M_i$ as for usual
enveloping algebras. It was also shown to be (non trivially)
isomorphic to a much more complicated `$\kappa$-Poincar\'e
algebra' which has obtained by contraction from the
$U_q(so_{3,2})$ $q$-deformation quantum group (see Section~III) but
for which no covariant Minkowski space had been known (until
\cite{MaRue:bic} only inconsistent non-covariant actions of this
quantum group on classical commutative Minkowski coordinates had
been considered). This completes the mathematics behind the
$\gamma$-ray bursts prediction of Section~II.C.

More recently, Connes and H. Moscovici studied a quantum group of
bicrossproduct form corresponding to the factorisation
\eqn{difffact}{ {\rm Diff}(\R)={\rm Diff}_0(\R)\cdot \R\lcross\R}
of diffeomorphisms into affine transformations $\R\lcross\R$ and
diffeomorphisms fixing the origin and the tangent space at the
origin. Using the algebraic bicrossproduct construction above we
have a bicrossproduct quantum group \eqn{conbica}{
\C[\R\lcross\R]\bicross U({\rm diff}_0(\R))} as well as a
covariant action of it on the noncommutative spacetime
$U(\R\lcross\R)$. This covariant action as well as the entirely
algebraic version of the construction go a little beyond
\cite{ConMos:hop}. In fact, what is more relevant there is
actually the dual model \eqn{conbicb}{
U(\R\lcross\R)\cobicross\C[{\rm Diff}_0]} where the coordinate
algebra of the diffeomorphisms was described as a certain
polynomial algebra in a countable number of generators
$\delta_n$. Next, just as the basis $u^n,v^n$ in Section~II.D for
the noncommutative torus could be taken as a model for physics
with integer $n$ replaced by knots, Connes and
D.~Kreimer\cite{ConKre:hop} proposed a variant of the above with
$\delta_n$ replaced by a labelling by rooted trees as a
book-keeping device for overlapping divergences in the
renormalisation a general quantum field theory. This is certainly
an interesting direction for current research.

\section{Quantum groups and $q$-deformation}

In this section we want to turn to the other and in many ways more
famous class of quantum groups, the deformed enveloping algebras
$U_q(\cg)$ and their associated coordinate algebras $\C_q[G]$. The
simplest and most well-known example of this type is the quantum
group $U_q(su_2)$ with generators $H,X_\pm$ and relations and
coproduct
\eqn{Uqsu2a}{ [H,X_\pm]=\pm 2 X_\pm,\quad
[X_+,X_-]=\frac{q^H-q^{-H}}{q-q^{-1}}}
\eqn{Uqsu2b}{ \Delta X_\pm=X_\pm\tens q^\frac{H}{2}+
q^\frac{-H}{2}\tens
X_\pm,\quad \Delta H=H\tens 1+1\tens H.} The coproduct $\Delta$
here is a deformation of the usual additive one, which is
recovered as $q\to 1$. The deformation modifies how an action of
$X_\pm$ extends to tensor products.

Likewise, the coordinate algebras $\C_q[G]$ are deformations of
the classical coordinate algebras $\C[G]$. As such one could say
{\em mathematically} that they are `quantisations' of the latter.
This is not quantum physics, however, because {\em any}
noncommutative algebra which is a nice deformation (a flat one) of
a commutative coordinate algebra of classical space implies (by
looking at the deformation to lowest order) a Poisson structure in
the classical space of which the noncommutative algebra could be
viewed as `quantisation'; this is tautological and it does not
necessarily come from these quantum groups arising as the algebra
of observables of any natural quantum system (in contrast to the
bicrossproduct quantum groups). Rather, the $U_q(\cg)$ have their
structure dictated by their roles as generalised symmetries. Their
key property is that their representations form a braided category
and that anything on which they act acquires braid statistics.
Afterwards one may indeed semiclassicalise and obtain new ideas
for Poisson geometry.

This is more or less but not quite what happened. On the one hand
quantum groups arose as symmetries in exactly solvable lattice
models and provided a new interpretation of the `corner transfer
matrix method' that had been developed by Baxter in the 1970s. On
the other hand there had independently been developing,
particularly in the then-Soviet Union, a theory of classical
integrable systems and `classical inverse scattering' for soliton
equations; see \cite{FadTak:ham} for a review. Initially this was
done through a notion of Lax pairs and zero curvature equations
but at the same time it turned out that many such systems could be
formulated in terms of a `classical Yang-Baxter equation' that one
obtains by semiclassicalising the methods for exactly solvable
lattice models. So quantum groups indeed turned out to form a
bridge between these schools, and it could be said that all three
influenced each other in their development.

Nevertheless the role of inducing braid statistics would appear to
be deeper than the `quantisation' point of view here and therefore
I will concentrate on this. It leads ultimately to a whole new
kind of `braided geometry' somewhat different from the
noncommutative geometry in the preceding section. As explained in
the introduction, the true meaning of the parameter $q$ in this
context is that it controls the braiding matrix that generalises
the minus sign in bose-fermi statistics to `braid statistics'.
This is the minus sign that is the origin of the Pauli exclusion
principle that two electrons cannot be in exactly the same state.
Note that the latter has a vague similarity with the Heisenberg
uncertainty principle itself; in fact the idea of noncommutativity
as in quantisation and the idea of braid statistics are intimately
related both mathematically and physically.

\subsection{Knot invariants}

Thus, if $V,W$ are representations of $U_q(\cg)$ then $V\tens W$
is (as for any quantum group) also a representation. The action is
\eqn{hact}{ h\la(v\tens w)=(\Delta h).(v\tens w)} for all $h\in
U_q(\cg)$, where we use the coproduct (for example the linear
form of the coproduct of $H$ means that it acts additively). The
special feature of quantum groups like $U_q(\cg)$ is that there
is an element $\CR\in U_q(\cg)\bar{\tens} U_q(\cg)$ (the
`universal R-matrix or quasitriangular structure') which ensures
an isomorphism of representations \eqn{Psi}{ \Psi_{V,W}:V\tens
W\to W\tens V,\quad \Psi_{V,W}(v\tens w)=P\circ \CR.(v\tens w)}
where $P$ is the usual permutation or flip map. This {\em
braiding} $\Psi$ behaves much like the usual transposition or
flip map for vector spaces but does not square to one. To reflect
this one writes $\Psi=\epsfbox{braid.eps}$,
$\Psi^{-1}=\epsfbox{braidinv.eps}$. It has properties consistent
with the braid relations, i.e. when two braids coincide the
compositions of $\Psi,\Psi^{-1}$ that they represent also
coincide. The fundamental braid relation of the braid group
corresponds to the Yang-Baxter or braid relation for the braiding
$\Psi$. This is shown in Figure~3 in terms of the matrix $R$
associated to $\Psi$ on choosing bases for the vector spaces.
This `R-matrix' is just the image of $\CR$ in the relevant
representations.
\begin{figure}
\[ R^i{}_a{}^k{}_b R^a{}_j{}^m{}_c R^b{}_l{}^c{}_n=
R^k{}_b{}^m{}_c R^i{}_a{}^c{}_n R^a{}_j{}^b{}_l\qquad
{\phantom{.}\atop\epsfbox{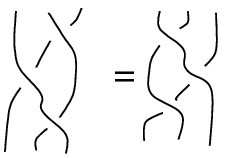}}\]
\caption{Braid relation and corresponding Yang-Baxter equation
for braiding $\Psi$ as a matrix $R$.}
\end{figure}

From this one can see how such quantum groups lead to knot
invariants\cite{ResTur:rib}. Thus, consider the knot as describing
the trajectories of particles $V$ and antiparticles $V^*$ with
time flowing down the page as in Figure~4(b). Note that rather
than thinking of a single kind of particle moving along the knot
(as in Figure~4(a)) we instead regard the upward arcs as
antiparticles flowing down the page. When one particle passes
over or under another, we apply some kind of operation $R$
according to the flavour of the crossing. In this way we `scan'
the knot from top to bottom, creating particles as needed,
interacting them at the crossings and finally fusing particles
and antiparticles as needed. The total process is computable and,
very roughly speaking, is the knot invariant as a function of any
parameter $q$ on which the matrices $R$ might depend.

\begin{figure}
\[ \epsfbox{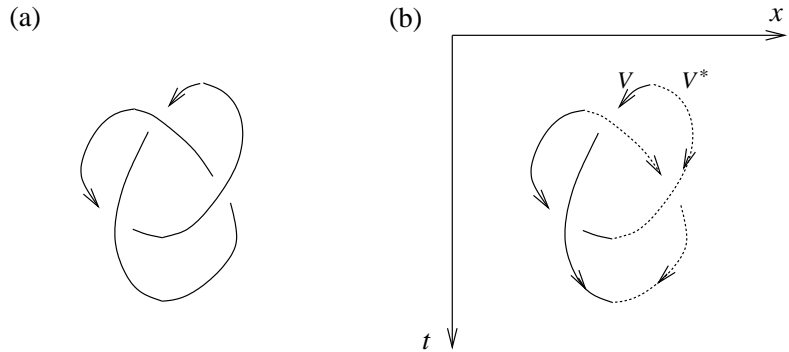}\]
\caption{Trefoil knot (a) and construction (b) of its invariant}
\end{figure}

Note that this process exists not in three dimensions (where our
original knot lived) but for particles and antiparticles moving in
one space and one time dimension, according to how the knot looks
on the page. We need to know that if we drew the knot from a
different angle and did our process from that point of view, we
should get the same answer. We also need to know that if we
distort the knot without cutting it then we get the same answer,
all of which depends on choosing $R$ carefully. The latter part of
the problem can be reduced mainly to the braid relation in
Figure~3. These braids are topologically the same so replacing one
by the other in a complicated knot would not change it. Therefore
we require that the corresponding operations $R$ should give the
same total process on three particles, which is the Yang-Baxter
equation. We also require similar relations where some strands are
antiparticles. The remainder of the problem can be focused mainly
on the observation that a harmless twist in the knot can appear
untwisted when viewed from a different angle, so that the number
of crossings themselves can change. In typical examples, the
matrices $R$, while not invariant under such harmless twists,
usually change in a simple way that can be compensated for by
hand. Actually, what one obtains in this way is not exactly a knot
invariant but an invariant of ribbons or framed knots. Apart from
these subtleties, these are the main constraints that the matrices
$R$ have to satisfy.

Also, in particle physics one understands particles as labelled by
representations $V$ of a Lie algebra $\cg$, and their conjugates
by the dual representation $V^*$. When particles are interchanged
one usually has either an exchange factor $R=-1$ (for Fermionic
particles like the electron) or the trivial exchange $R=1$ (for
Bosonic particles like the photon). Neither of these choices give
interesting knot invariants, but when we look instead at
representations of the quantum groups $U_q(\cg)$ we find a much
more nontrivial matrix $R$ (depending on $q$) whenever two
representations of the quantum group are exchanged. We can then
proceed as with the heuristic particle picture above but with $V$
a representation of the quantum group $U_q(\cg)$ and $V^*$ its
dual representation. They are created together as the canonical
element of $V\tens V^*$ (or a certain other element of $V^*\tens
V$) and are fused by the evaluation map $V^*\tens V\to
\C$ (or a certain other map $V\tens V^*\to\C$). The result is a
function of $q$ and this, more precisely, is the construction of
the knot invariants from quantum groups.

For standard $U_q(\cg)$ and generic $q$ the construction of
representations is not hard, all the standard ones of $\cg$ just
$q$-deform. For example, the spin $\h$ representation of $su_2$
deforms to a 2-dimensional representation of $U_q(su_2)$. The
associated knot invariant is the Jones polynomial knot
invariant\cite{Jon:pol}. Jones himself came to this more from the
solvable lattice point of view (see below), with the quantum
groups point of view coming later.

There is a lot more to the deep mathematical structure of the
$U_q(\cg)$ that is not that much applied so far in physics. In
particular, there is an important subalgebra $U_q(n_-)\subseteq
U_q(\cg)$ which can be used to generate representations of the
whole quantum group from a highest-weight `vacuum' vector.
G.~Lusztig in the early 1990s introduced a nice description of
this subalgebra in terms of certain (shifted) `perverse sheaves',
using concepts from algebraic geometry \cite{Lus}. Without going
into details, he obtained in this way a basis of the subalgebra
with many remarkable integrality and positivity properties, called
the {\em Kashiwara-Lusztig canonical basis} (a similar `global
crystal basis' was found by Kashiwara at about the same time). The
most remarkable property of this basis is that it induces a basis
of every highest weight representation that $U_q(n_-)$ generates.
This might seem esoteric but all these results continue to hold
even when $q=1$, and as such they provided unsuspected and
 revolutionary results in the representation theory of ordinary
Lie algebras $\cg$ themselves. The algebras of negative roots
$U_q(n_-)$ themselves are not actually quantum groups but braided
versions or `braided groups' as introduced by the author (see a
later subsection). Using some of the theorems for these objects,
one has further results such as an inductive construction of the
$U_q(\cg)$ as a decomposition into a series of quantum-braided
planes\cite{Ma:dbos}\cite{Ma:lieb}. Again, the result is useful
even for $U(\cg)$ and is relevant to their
noncommutative-geometric picture as in Section~II.

\subsection{Exactly solvable lattice models}

Here we give the briefest of expositions of how, at least
conceptually, quantum groups and braiding matrices $R$ came out of
solid state physics in the early 1980s. This is surely one of the
great triumphs of the interaction between physics and pure
mathematics.

We recall that in statistical mechanics one has a large collection
of distinct states of the system and studies its bulk properties
through the partition function, a certain weighted sum over the
states. For example, consider the model of a crystal in Figure~5,
where a state is an assignment of bonds throughout the lattice. We
write the {\em Boltzmann weight} at each vertex as the entry of a
matrix $R$ according to the value of the bonds around the vertex.
The partition function is
\eqn{Partn}{ Z(\lambda)=\sum_{{\rm states}}\prod_{{\rm vertices}}
R^i{}_j{}^k{}_l(\lambda),} where $ijkl$ are the values of the
bonds in the given state surrounding the given vertex. We suppose
the weight depends on a parameter $\lambda$.
\begin{figure}
\[ R^i{}_j{}^k{}_l(\lambda)\qquad { \phantom{.}
\atop\epsfbox{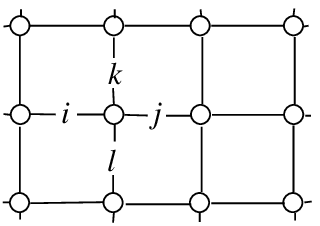}}\]
\caption{Solvable lattice model in statistical
mechanics has weight $R$ at each vertex}
\end{figure}
Working in a different (but broadly equivalent) setting, R. Baxter
in the early 1970s described conditions on $R$ which allowed for
the partition function to be computed explicitly using a `corner
transfer matrix method'\cite{Bax:exa}. The resulting functions
often had beautiful connections with number theory and the theory
of modular forms (not connected with A. Wiles' recent proof of
Fermat's theorem but in the same general ball-park). The required
conditions were that $R$ depends on a parameter $\lambda$ and
obeys a parameterised version of the Yang-Baxter equation in
Figure~3. So the key idea behind the knot invariants also makes
these models solvable. Later on, I. Sklyanin, L.D. Faddeev and
others recast the corner transfer matrix method more algebraically
in terms of an abstract algebra with generators $t^i{}_j(\lambda)$
 and relations
\eqn{RTTL}{R^i{}_a{}^k{}_b(\lambda+\mu)
t^a{}_j(\lambda)t^b{}_l(\mu)=t^k{}_b(\mu)t^i{}_a(\lambda)
R^a{}_j{}^b{}_l(\lambda+\mu).}
Using the Yang-Baxter equations many times it is easy to see that
the product of the weights along an entire row of the lattice and
with $i,j$ on the ends of the row,
\eqn{singlerow}{ (t^i{}_j){}^K{}_L=R^i{}_{a_1}{}^{k_1}{}_{l_1}
R^{a_1}{}_{a_2}{}^{k_2}{}_{l_2}\cdots
R^{a_{N-1}}{}_j{}^{k_N}{}_{l_N}} is a matrix representation of the
above algebra. Here there are $N$ columns, say, and
$K=(k_1,\cdots,k_N)$ etc. is a multiindex. If we are interested in
periodic boundary conditions then we should consider the trace
$T(\lambda)=t^i{}_i(\lambda)$. On the other hand for any operators
obeying the relations (\ref{RTTL}) it is easy to see that
\eqn{tracecomm}{ [T(\lambda),T(\mu)]=0}
for all $\lambda,\mu$ i.e. the single row transfer matrices
$T(\lambda)$ form an infinite number of mutually commuting
operators. This is the origin of the exact solvability of the
models. The partition function itself is the trace over the
multiindex of the $T(\lambda)$. Another variant of this method
based on open or twisted boundary conditions involved instead a
`reflection' form of (\ref{RTTL}) where two $R$'s appear on both
sides.

One does not actually need the abstract algebra here -- it is just
a convenient way to encode certain computations involving repeated
use of the Yang-Baxter equations. On the other hand, there is a
kind of formal coproduct structure \eqn{deltaTL}{ \Delta
t^i{}_j(\lambda)=t^i{}_a(\lambda) \tens t^a{}_j(\lambda)}
(forming a quantum group without antipode) which is the underlying
reason that (\ref{singlerow}) is a representation, namely it is
the repeated tensor product of the 1-column representation
$(t^i{}_j)^k{}_l=R^i{}_j{}^k{}_l$ provided by $R$ itself (which is
easier to see -- it is just the Yang-Baxter equation). This
observation is not true for the other `reflection' variant (which
tends to be more like a braided group as we will see in a later
section). More importantly, the focus on an abstract algebra
(\ref{RTTL}) suggested the possibility to realise such algebras in
terms of other simpler algebras by means of various ansatze. One
of the models was the so-called XXZ model consisting of
nearest-neighbour spin interactions and a uniform magnetic field
(controlled by a parameter $q$) running through the lattice, and
this turned out to be realised in terms of the quantum group
$U_q(su_2)$. So this quantum group is in the background of this
model playing the role of inducing the required
$R(\lambda)$-matrices. One can also focus on the
finite-dimensional version of the above by looking at $R(\infty)$
or $R(0)$ etc. which takes us to the simpler $R$-matrices used
above for the construction of knots.

We should also consider the continuum limits of such models as the
lattice spacing tends to zero. In many cases one obtains a
conformally-invariant quantum field theory. Such `conformal field
theories' turned out to have their own rich algebraic structure of
vertex algebras and were connected with modular forms, the
`monster group' and other topics. One of them (the
Wess-Zumino-Novikov-Witten model) underlies the quantum group knot
invariants above. And the classical mechanical systems underlying
the continuum limits of the exactly solvable lattice models turned
out to be a certain non-linear but completely integrable partial
differential equations with `soliton' solutions. So one can trace
a certain continuity of ideas through several key developments in
mathematical physics.

\subsection{$q$-coordinate algebras and Poisson-Lie groups}

If one did want to take a quantisation point of view then one
should consider not so much $U_q(\cg)$ as we have done above but
their coordinate algebra $\C_q[G]$. These are also suggested by
the lattice model picture.

We consider first of all the {\em quantum plane} $\C_q[x,y]$. This
is the algebra generated by variables $x,y$ but with the relations
\eqn{qplane}{yx=qxy}
instead of commutativity. Here $q$ is a non-zero numerical
parameter. When $q=1$ we can consider $x,y$ as the coordinates on
an actual plane as we did above, but when $q\ne 1$ the algebra is
noncommutative and hence there is no usual space underlying it. We
also have higher-dimensional quantum spaces of many kinds
depending on the relations and parameters. In particular, the
quantum group $\C_q[SU_2]$ has generators $a,b,c,d$ with the six
relations
\[ ba=qab,\quad dc=qcd,\quad ca=qac,\quad db=qbd\]
\eqn{qmat}{ bc=cb,\quad ad-da=(q^{-1}-q)bc}
which describe a 4-dimensional $q$-space (they become the
relations of commutativity when $q=1$), and the additional
relation
\eqn{qdet}{ ad-q^{-1}bc=1}
which sets the `$q$-determinant' to $1$. There is also a $*$
operation to express unitarity. There is nothing much that need
concern us about the exact form of the above relations; the main
thing is that as $q=1$ they recover the commutativity and
determinant relations that we expect for the coordinates on the
classical group $SU_2$ of $2\times 2$ matrices of determinant 1.
Their exact form is, however, fine-tuned in such a way that
various properties of $2\times 2$ matrices and their action on
vectors go through even when $q\ne 1$. Thus, if $x,y$ generate a
quantum plane then
\eqn{trans}{ x'=ax+by,\quad y'=cx+dy}
obey the relations $y'x'=qx'y'$ of the quantum plane as well. In
mathematical terms this `quantum transformation' is an algebra map
$\Delta_L:\C_q[x,y]\to \C_q[SU_2]\tens \C_q[x,y]$ called a
`coaction'. Note that the arrow goes in the reverse direction to
what one might have expected if one thought that an actual matrix
was being combined with a vector to give another vector.

To complete the picture here, we need to check that the group
structure itself is expressed in our algebraic language. In the
above example, the ability to multiply two matrices to get a third
matrix corresponds to the assertion that if $a',b',c',d'$ are a
second mutually commuting copy of $\C_q[SU_2]$ then
\eqn{matmul}{ \begin{pmatrix}
 a'' & b''\\ c''& d''\end{pmatrix}=\begin{pmatrix}a & b\\ c&
 d\end{pmatrix}\begin{pmatrix}a' & b'\\ c'& d'\end{pmatrix}} obeys
the same relations. In mathematical terms the group law is
expressed as an algebra map $\Delta:\C_q[SU_2]\to \C_q[SU_2]\tens
\C_q[SU_2]$ which is the coproduct of the quantum group. It has
the same matrix form $\Delta a=a\tens a+b\tens c$ etc., as for
$\C[SU_2]$ in Section~II.C. These constructions are algebraic but
one can cast them into an operator algebra setting as in
Section~II\cite{Wor:twi}.

Next we note that in the above example and many like it the
$q$-commutativity relations can be cast as \eqn{RTT}{
R^i{}_a{}^k{}_b t^a{}_j t^b{}_l=t^k{}_b t^i{}_a R^a{}_j{}^b{}_l}
for suitable $R$ obeying the Yang-Baxter equations. This is a
`constant' version\cite{FRT:lie} of (\ref{RTTL}) focussed on
$\lambda=\infty$ or $\lambda=0$ etc. (one has to make some other
equivalences to get to the actual form of $R$ from the more
physical $R(\lambda)$ and one should also note that in
Section~III.B the generators are more related to enveloping
algebras than to coordinate algebras). Initially it was often
mistakenly written that the Yang-Baxter relation is what makes
these algebras $A(R)$ into quantum groups, which is not at all
true. In fact for any tensor $c_{i_1\cdots i_n}{}^{j_1\cdots
j_m}$ there is a quantum group (without antipode) $M(c)$ with a
matrix $t^i{}_j$ of generators and the relations \eqn{auto}{
c_{a_1\cdots a_n}{}^{j_1\cdots j_m}t^{a_1}{}_{i_1} \cdots
t^{a_n}{}_{i_n} = t^{j_1}{}_{b_1}\cdots
t^{j_m}{}_{b_m}c_{i_1\cdots i_n}{}^{b_1 \cdots b_m}.} For
example, $c^i{}_{jk}$ could be the structure constants of an
associative algebra and then $M(c)$ is its universal comeasuring
(or `automorphism') quantum group\cite{Ma:dif}. Rather, the
meaning of $R$ obeying the Yang-Baxter equations is that this
ensures that the comodules of $A(R)$ form a braided category,
i.e. that quantum groups such as $\C_q[SU_2]$ are quasitriangular
in a comodule sense.

Probably the most immediate significance of these quantum group
coordinate algebras $\C_q[G]$ is that one can consider them
formally as `quantisations' of interesting Poisson brackets on the
classical group $G$ (even though they are not really the algebra
of observables of a true quantum system and $q$ need not be
related to Planck's constant). The Poisson brackets so obtained by
semiclassicalisation are always degenerate at the group identity,
so this kind of Poisson bracket was missed by those focusing on
symplectic manifolds only. Instead they form a {\em Poisson-Lie
group} in the sense that the group product $G\times G\to G$
respects the Poisson structure (taking the direct product Poisson
structure on $G\times G$). Among the matrix $t^i{}_j$ of
coordinates in $\C[G]$ the Poisson bracket has the form
\eqn{poisl}{ \{t^i{}_j,t^k{}_l\}=t^i{}_a t^k{}_b r^a{}_j{}^b{}_l
-r^i{}_a{}^k{}_b t^a{}_j t^b{}_l}
where $r$ is the lowest order deviation from the identity matrix
of $R$. Clearly $r$ obeys an infinitesimal version of the
Yang-Baxter equation, called the `classical Yang-Baxter equation'.
The abstract picture here is perhaps more easily seen in a dual
form as the lowest order part in the deformation from $U(\cg)$ to
$U_q(\cg)$. Thus the deformation of the coproduct to lowest order
is a `Lie cobracket' $\delta:\cg\to\cg\tens\cg$ forming a Lie
coalgebra (so that $\cg^*$ is a Lie algebra) and respecting the
Lie bracket of $\cg$ in a suitable sense. This is the
infinitesimal analogue of a quantum group and is called a Lie
bialgebra\cite{Dri:ham}. Such a $\delta$ typically extends to all
of $G$ as a bivector field which defines the Poisson bracket. In
the quasitriangular case as above it has the special form
$\delta\xi={\rm ad}_\xi(r)$ where $r\in \cg\tens \cg$ obeys the
abstract classical Yang-Baxter equation
\eqn{CYBE}{ [r_{12},r_{13}]+[r_{12},r_{23}]+[r_{13},r_{23}]=0.}
The numerical suffices here denote in which factor of
$\cg\tens\cg\tens\cg$ one should view the two legs of $r$. The
latter is the leading deviation from $1$ of the quasitriangular
structure of $U_q(\cg)$ mentioned in Section~III.A.

These ideas have allowed mathematicians to go back and understand
many constructions in conventional Lie theory in a more elegant
and natural manner, as well as to obtain entirely new results.
They also allow one to present a cleaner treatment (at the level
of finite-dimensional Lie algebras) of the integrability of the
classical mechanical systems underlying the solvable lattice
models of Section~III.B. At least at the simplified level an
outline is as follows. Suppose that a Lie group $X$ factorises
into $GM$ and its Lie algebra $\cx$ has a nondegenerate
ad-invariant bilinear form $K$. Using the latter one can view the
difference of the projection operators $\pi_\pm$ on $\cx$
corresponding to $\cx=\cg\oplus\cm$ as defining a solution
$r\in\cx\tens \cx$. This equips $X$ with a certain Poisson
bracket. Now given a choice of Ad-invariant Hamiltonian function
$h$ on $X$ one may analyse the induced Hamilton-Jacobi equations
of motion in terms of $G,M$ and find that they are completely
solvable. In fact the evolution of $x\in X$ is given by the
Adler-Kostant-Symes theorem \eqn{AKSa}{ x_t=s_t
x_0s_t^{-1}=u_t^{-1}x_0u_t,} where $u_t,s_t$ are paths in $G,M$
determined by the factorisation \eqn{AKSb}{ e^{-t K^{-1}\circ
L_{x_0*}^*(\extd h)}=u_t s_t} of an exponential flow in $X$. Here
$L_{x*}$ denotes the differential of left multiplication in $X$.
For the actual nonlinear integrable systems of interest with
solitons etc., one should work with a parameterised version of
similar constructions. In this case the factorisation is
typically that of loop groups into loops in a target group that
are analytic outside and inside the unit disc in the parameter
space, which is the classic Riemann-Hilbert factorisation
problem. We refer to \cite{ReySem:redII} for a fuller treatment
and the relation with Lax pairs and other topics. This makes a
little more precise the remarks at the end of Section~III.B.

\subsection{Braided geometry and $q$-spacetime}

After the great success of the quantum groups $U_q(\cg)$ there was
a period in the 1990s when physicists enthusiastically went about
$q$-deforming everything they could think of where Lie groups has
been involved. One has
$q$-oscillators\cite{Mac:ana}\cite{Bie:har}, $q$-Brownian motion,
etc. etc. It is not clear what it all adds up to in the longer
term but there clearly does appear to be a natural (if not exactly
unique) $q$-deformation of almost everything.

And according to what we have said in the introduction, the
systematic way to go about doing this was {\em braided geometry}.
This is because the braiding is the key property of the quantum
groups $U_q(\cg)$ and other `quasitriangular Hopf algebras' of
similar type. It meant in particular that any algebra on which the
quantum group acts covariantly becomes braided, which was
therefore indicative of a whole braided approach to noncommutative
geometry via algebras or `braided' spaces on which quantum groups
$U_q(\cg)$ act as generalised symmetries. Note that we are not so
much interested from this point of view in the noncommutative
geometry of the quantum groups $U_q(\cg)$ themselves, although one
can study this as a source of mathematical examples. This
systematic braided approach was introduced by the author at the
end of the 1980s; see \cite{Ma:introp}\cite{Ma:varen} for reviews.

To get an immediate flavour for what these ideas mean in practice,
consider the following elementary computation. For a polynomial
function $f$ in one variable, define differentiation by
\eqn{qdifa}{ f'(y)=(x^{-1}(f(x+y)-f(y)))_{x=0}.} If $xy=yx$ is
assumed in making the calculation, one obtains the usual
Newtonian differentiation. But if we suppose $yx=qxy$ in
computing the right hand side, for some parameter $q$, we obtain
\eqn{qdif}{ f'(x)=\frac{f(x)-f(qx)}{ (1-q)x}.} This is the
celebrated `$q$-deformed derivative', so called because it tends
to the usual derivative as $q\to 1$. Although known to
mathematicians in a different context in 1908\cite{Jac:fun}, such
$q$-derivatives have their natural place in the geometry of
quantum groups. We also see by this example that noncommutativity
leads to a kind of `finite-difference' or discretisation, which
is therefore a general feature of the differential geometry of
the quantum world and also in keeping with applications to
lattice models. This point of view also leads to the correct
properties of integration. Namely there is a relevant indefinite
integration to go with $\del_q$ characterised by\cite{KemMa:alg}
\eqn{$q$-int}{\int_0^{x+y}f=\int_0^x f((\ )+y)+\int_0^y f}
provided $yx=qxy$, etc. In the limit this gives the infinite
Jackson integral previously known in this context. One also has
braided exponentials, braided Fourier theory etc., for these
braided variables. What is going on here is that one is working
not with the usual line with bosonic coordinate $x$ but with the
{\em braided line}. This is the braided group $B=\C[x]$, which is
the usual polynomial algebra but with braid statistics, such that
two copies of the braided line have the relations $yx=qxy$. The
addition $x+y$ refers to an additive braided group structure or
coproduct.

This is a powerful point of view and more systematic than simply
sprinkling in $q$ by trial and error. For example, we have
$B=\C_q[x,y]$ the quantum-braided plane generated by $x,y$ with
the relations $yx=qxy$, where two independent copies have the
braid statistics \eqn{brapla}{ x'x=q^2xx',\quad x'y=qyx',\quad
y'y=q^2yy',\quad y'x=qxy'+(q^2-1)yx'.} Here $x',y'$ are the
generators of the second copy of the plane. There is again an
additive braided coproduct in the sense that $x+x',y+y'$ is
another copy of the quantum-braided plane, i.e. one can check that
\eqn{addpla}{ (y+y')(x+x')=q(x+x')(y+y').} And by similar
definitions as above, one has braided partial derivatives
\eqn{delpla}{ \del_{q,x}f(x,y)=\frac{f(x,y)-f(qx,y)}{(1-q)x},\quad
\del_{q,y}f(x,y)=\frac{f(qx,y)-f(qx,qy)}{(1-q)y}} for expressions
normal ordered to $x$ on the left, etc. Note in the second
expression an extra $q$ as $\del_{q,y}$ moves past the $x$ to act
on $y$.

Thus {\em you can add points in the quantum-braided plane}, and
then (by an infinitesimal addition) define partial derivatives
etc. One then has multivariable $q$-exponentials and so on. This
is a problem (multivariable $q$-analysis) which had been open
since 1908 and was systematically solved in the early 1990s by the
braided approach \cite{Ma:poi}. We note in passing that $yx=qxy$
is sometimes called the `Manin plane'. Manin considered only the
algebra and a quantum group action on it, without the braided
addition law and the braided approach.

Finally, there is a more formal way by which all such
constructions are done systematically, which we now explain. It
amounts to nothing less than a new kind of algebra in which
algebraic symbols are replaced by braids and knots. First of all,
given two algebras $B,C$ in a braided category (such as the
representation of $U_q(\cg)$) we have a braided tensor product
$B\und\tens C$ algebra in the same category defined like a
superalgebra but with $-1$ replaced by the braiding
$\Psi_{C,B}:C\tens B\to B\tens C$. Thus the tensor product becomes
noncommutative (even if each algebra $B,C$ was commutative) -- the
two subalgebras `commute' up to $\Psi$. This is the mathematical
definition of braid statistics. In the braided line the joint
algebra of the independent $x,y$ is $\C[x]\und\tens\C[y]$ with
$\Psi(x\tens y)=qy\tens x$. In the braided plane the braided
tensor product is between one copy $x,y$ and the other $x',y'$.
The braiding $\Psi$ in this case is more complicated. In fact it
is the same braiding from the $U_q(su_2)$ spin $\h$ representation
that gave the Jones polynomial. The miracle that makes knot
invariants is the same miracle that allows braided multilinear
algebra and multivariable $q$-analysis.

The addition law in both the above examples makes them into
braided groups\cite{Ma:bg}. They are like quantum groups or
super-quantum groups but with braid statistics. Thus, there is a
coproduct
\eqn{coppla}{ \Delta x=x\tens 1+1\tens x,\quad \Delta y=y\tens 1
+1\tens y}
etc., (this is a more formal way to write $x+x',y+y'$). But
$\Delta:B\to B\und\tens B$ rather than mapping to the usual tensor
product. We then write the braiding $\Psi$ as a braid crossing for
reasons explained in Section~III.A. In this language one also draws
the product $B\tens B\to B$ as a map $\epsfbox{prodfrag.eps}$, the
coproduct as $\epsfbox{deltafrag.eps}$, etc. Similarly with other
maps, some strands coming in for the inputs and some leaving for
the outputs. We then `wire up' an algebraic expression by wiring
outputs of one operation into the inputs of others, flowing down
the page. When wires have to cross under or over, we have to chose
one or the other as $\Psi$ or $\Psi^{-1}$. Algebra is then done as
equalities of branched braids. For example, in Figure~6(a) we
write the associativity of the product of an algebra and the
homomorphism property of the coproduct $\Delta$ for a Hopf algebra
or `braided group' in this context. The coassociativity of the
coproduct is the first part of Figure~6(a) turned up side down.
There are also unity and antipode axioms. Part (b) of the figure
is a example of a braided algebra calculation.
\begin{figure}
\[ {\rm (a)}\quad\epsfbox{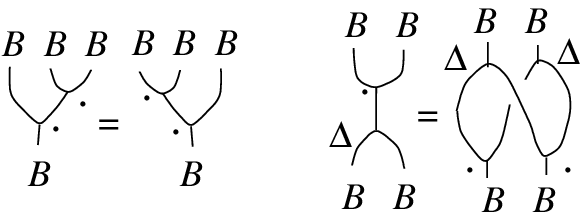}\quad\quad {\rm (b)}
\quad\epsfbox{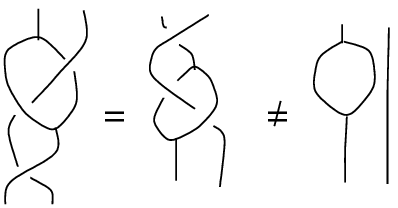}\]
\caption{(a) Main axioms of a braided group as diagrams and (b)
example of a braided calculation}
\end{figure}
Using this method one finds that all of the usual constructions
for groups and quantum groups go through. For example, the notion
of adjoint action or conjugation is shown in Figure~7. On the left
is a `breakdown' of the steps involved in usual conjugation in a
group. One can think of a group trivially as a braided group with
a coproduct that simply doubles up the group element. We double up
$h$ in this way, move one of these past the $g$, apply the
antipode or inversion operation $S$ and then multiply up. The
corresponding diagram is shown on the right. In this way one
arrives at a theory of algebras and groups which exists entirely
at the level of braids and branches. One can do proofs, roughly
speaking, by treating these as actual strings, i.e. this is a kind
of knot-theoretic algebra. Fourier theory etc. all go through at
this level.
\begin{figure}
\[ {\rm Ad}_h(g)={\phantom{.}\atop\epsfbox{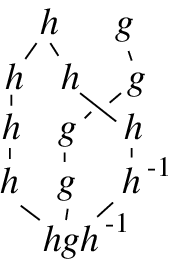}}=hgh^{-1}\quad
\quad {\rm Ad}={\phantom{.}\atop\epsfbox{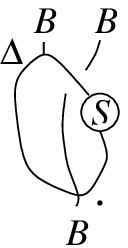}}\]
\caption{Conjugation written as a diagram}
\end{figure}

Clearly such braided groups are, in particular, the correct
foundation for $q$-deformed geometry based on $q$-planes and
similar $q$-spaces. One of their main successes in the early 1990s
was a more or less complete and systematic $q$-deformation of the
main structures of special relativity and electromagnetism, i.e.
$q$-Minkowski space and basic structures,
\begin{itemize}
\item $q$-Minkowski space as $2\times 2$ braided Hermitian matrices

\item $q$-addition etc., on $q$-Minkowski space

\item $q$-Lorentz quantum group as a double cross product $\C_q[SU_2]
\dcross \C_q[SU_2]$

\item $q$-Poincar\'e+scale quantum group $\R_q^{1,3}\lbiprod
\widetilde{U_q(so_{1,3})}$

\item $q$-partial derivatives

\item $q$-differential forms

\item $q$-epsilon tensor

\item $q$-metric

\item $q$-integration with Gaussian weight

\item $q$-Fourier theory

\item $q$-Green functions (but no closed form)

\item $q$-$*$ structures and $q$-Wick rotation

\item $q$-conformal group $\R_q^{1,3}\lbiprod
\widetilde{U_q(so_{1,3})}\rbiprod\R_q^{1,3}$

\item $q$-diffeomorphism groups
\end{itemize}

This general braided approach
\cite{Ma:exa}\cite{MaMey:bra}\cite{Ma:poi}\cite{Ma:fre}\cite{Ma:eps}
\cite{KemMa:alg}\cite{Ma:star}\cite{Ma:qsta}\cite{Ma:euc}
\cite{Ma:conf}\cite{Ma:dif} worked for any braiding or
`R-matrix'. For example, the correct notion of `braided matrices'
$B(R)$ that goes along with the braided plane etc. above is
generated by $u^i{}_j$ with the braided matrix relations
\eqn{B(R)}{ R^k{}_a{}^i{}_b u^b{}_c R^c{}_j{}^a{}_d u^d{}_l
=u^k{}_a R^a{}_b{}^i{}_c u^c{}_d R^d{}_j{}^b{}_l.} Just as the
quantum matrices $A(R)$ in the previous section could be viewed
as parameter-free versions of the algebra (\ref{RTTL}) but have
their own life as geometric objects, these braided matrices
$B(R)$ could also be viewed as a parameter-free version of the
variants of (\ref{RTTL}) with two $R$'s on each side. On the
other hand all of their key properties, such as covariance under
a background quantum group, the braid statistics and braided
coproduct \eqn{bramult}{ R^{-1}{}^i{}_b{}^k{}_au'{}^b{}_c
R^c{}_j{}^a{}_du^d{}_l =u^k{}_a R^{-1}{}^i{}_c{}^a{}_b u'{}^c{}_d
R^d{}_j{}^b{}_l,\quad \Delta u^i{}_j=u^i{}_a\tens u^a{}_j} and so
on, came out of the theory of braided groups as part of a close
mathematical relationship called {\em transmutation} between
$A(R)$ and $B(R)$ i.e. from the quantum to the braided versions.
In fact this suggests a physical equivalence between the periodic
and the open lattice systems by such transmutation that has not
been explored much so far.

The specific choice of the same R-matrix as for the quantum plane
or for $\C_q[SU_2]$ gives the algebra\cite{Ma:exa}
\[ba=q^2ab,\quad ca=q^{-2}ac,\quad da=ad,\quad bc=cb
+(1-q^{-2})a(d-a),\] \eqn{qmink}{db=bd+(1-q^{-2})ab,\qquad
cd=dc+(1-q^{-2})ca} of $q$-Minkowski space. This is the version
with a central time $t=q^{-1}a+qd$ as promised. Similarly one can
plug the standard R-matrix into general braided constructions and
obtain the specific form of the other required $q$-algebras. In
some cases this braided approach was confirmed by recovering
algebras which had been proposed independently for this case by
more ad hoc methods. In particular, the relations (\ref{qmink})
were proposed as $q$-Minkowski space in \cite{CWSSW:ten} as an
algebra on which the $q$-Lorentz group acts. Similarly Wess,
Zumino et al. first proposed a large algebra with many generators
and $q$-relations as $q$-Poincar\'e in \cite{OSWZ:def}. The
braided approach provided R-matrix formulae, the braided addition
and braided matrix structures of $q$-Minkowski space and the
covariant action of $q$-Poincar\'e it. The latter came out of a
general {\em bosonisation} theorem which asserts that for every
braided group $B$ with background symmetry a (co)quasitriangular
quantum group $H$ one has an equivalent ordinary (bosonic)
quantum group $B\lbiprod H$. The main thing is that from this
communal effort a more or less definitive version of the
mathematical structures for this particular toy model of
spacetime is now known. In particular, one has $q$-Fourier
transform to $q$-momentum space which turns out to be
$q$-Minkowski space again (just like for $\R^n$). In the language
of Section~II it means that there are both `quantum' and `curved'
aspects matching each other and isomorphic via the quantum
metric. Using this one has no problem to define formal
powerseries for the $q$-Green function as the inverse Fourier
transform  \eqn{$q$-green}{ G_q(\vec x)=\CF^{-1}(\vec p\cdot \vec
p-m^2)^{-1}} so that, in principle, this is now defined.

There are some fundamental problems at the moment with this
$q$-spacetime before one can expect real physical predictions (in
contrast to the simpler model in Section~II). First of all one does
not generally have closed expressions such as for the $q$-Green
functions above. The methods of $q$-analysis as in
\cite{Jac:fun}\cite{Koo:ort} are simply not yet far enough
advanced to have nice names and properties for the kinds of
powerseries functions encountered. This is a matter of time.
Secondly, while the $q$-Poincar\'e coordinate algebra has a
natural $*$-algebra structure so that one can study its
representations in Hilbert spaces etc. (this was first done in the
Euclidean case by G.~Fiore) the $*$-structure does not respect the
coproduct in the obvious way. One can understand this as rather
fundamental to braided geometry for the following reason: when we
deform classical constructions to braided ones we have to choose
$\Psi$ or $\Psi^{-1}$ whenever wires cross. Sometimes neither will
do, things get tangled up. But if we succeed it means that for
every $q$-deformation there is another where we could have made
the opposite choice in every case. So classical geometry
bifurcates into two $q$-deformed geometries according to $\Psi$ or
$\Psi^{-1}$. Moreover, the role of the $*$ operation is that it
interchanges these two\cite{Ma:qsta}. Roughly speaking,
\eqn{splitgeom}{ \begin{matrix}
& \nearrow& $q$-{\rm geometry}\\ {\rm classical\ geometry}&
&\updownarrow *\\ & \searrow& {\rm conjugate}\ $q$-{\rm geometry}
\end{matrix}  }
where the conjugate is constructed by interchanging the braiding
with the inverse braiding (i.e. reversing braid crossings in the
diagrammatic construction). For the simplest cases like the
braided line it means interchanging $q,q^{-1}$. This is rather
interesting given that the $*$-operation is a central foundation
of quantum mechanics and our concepts of probability. But it also
means one probably cannot do $q$-quantum mechanics etc., with
$q$-geometry alone; one needs also the conjugate geometry.

Although there are such difficulties which make it hard at the
moment to assess the physical significance of this kind of
$q$-deformation, there are some important motivations and one
should expect that some of them will eventually be realised using
some later version of our present efforts in this direction. First
of all we have said that the true meaning of $q$ is that it
generalises the $-1$ of fermionic statistics. That is why it is
dimensionless. It is nothing other than a parameter in a
mathematical structure (the braiding) in a generalisation of our
usual concepts of algebra and geometry, going a step beyond
supergeometry. It is very likely, if not clear, that $q$ a root of
unity would therefore be the correct setting for the treatment of
certain anyonic systems where particles of anyonic statistics
should be found. Or conversely one should identify and study known
models with $q$-symmetry at $q$ a root of unity from the point of
view of identifying the modes with anyonic statistics. This is
clear but should be elaborated further. $q$-deformed constructions
should then help in understanding the geometry of such systems.
And of course there is the original physical meaning where $q$ is
related to an anisotropy such as that due to an external magnetic
field.

Other than these, there are some potential long-term reasons to
$q$-deform, particularly spacetime as above. Thus, in
\cite{Ma:reg} it was proposed that since $q$ was dimensionless and
somewhat `orthogonal' to physics it should be an ideal parameter
for regularising any quantum field theory. Since most
constructions in physics $q$-deform, such a regularisation scheme
is much less brutal than say dimensional or Pauli-Villars
regularisation as it preserves symmetries as $q$-symmetries, the
$q$-epsilon tensor etc. In this context it seems at first {\em too
good} a regularisation. Something has to go wrong for anomalies to
appear. For example, it would be interesting to see exactly how
the axial anomaly appears in this regularisation approach. The
main thing that the regularisation loses is that only the
Poincar\'e+scale $q$-deforms (the two get mixed up) which means
that only massless particles should be treated in the fist place.
The massive case would break even this invariance, giving
different results for the two. Pushing the problem into the scale
generator also suggests that a much nicer treatment of the
renormalisation group should be possible in this context. Again a
lot of this must await more development of the tools of
$q$-analysis. At any rate the result in \cite{Ma:reg} is that
$q$-deformation does indeed regularise, turning some of the
infinities from a Feynman loop integration into poles
$(q-1)^{-1}$.

Also, $q$-deformation might be useful as a next-order
approximation to the geometry coming out of a known or unknown
theory of quantum gravity. Thus, as well as being a good regulator
one can envisage (in view of our general ideas about
noncommutativity and the Planck scale) that the actual world is in
reality better described by $q\ne 1$ due to Planck scale effects.
This was the original reason given in \cite{Ma:reg} for
 $q$-deforming the basic structures of physics. The UV
cut-off provided by a `foam-like structure of space time' would
instead be provided by $q\ne 1$. Moreover, if this is so then
$q$-deformed quantum field theory should also appear coming out of
quantum gravity as an approximation one better than the usual.
Such a theory would be massless according to the above remarks
(because there is no $q$-Poincar\'e without the scale generator).
Or at least particle masses would be small compared to the Planck
mass. How the $q$-scale invariance breaks would then be a
mechanism for mass generation.

Recently, it was argued\cite{MajSmo:def} that since loop gravity
is linked to the Wess-Zumino-Novikov-Witten (WZNW) model, which is
linked to $U_q(su_2)$ (or some other quantum group), that indeed
$q$-geometry should appear coming out of quantum-gravity with
cosmological constant $\Lambda$. There is even provided a formula
\eqn{smolin}{ q=e^\frac{2\pi\imath}{2+k},\quad k=\frac{6\pi}
{G_{\rm Newton}^2\Lambda}.} If so then the many tools of
$q$-deformation developed in the 1990s would suddenly be
applicable to study the next-to-classical structure of
quantum-gravity. The fact that loop variable and spin-network
methods `tap into' the revolutions that have taken place in the
last decade around quantum groups, knot theory and the WZNW model
makes such a conjecture reasonable.

Whether it is a matter of $q$-regularisation of flat space or of
actual Planck scale effects, there are several {\em new} things
that happen for $q\ne 1$ which are not visible for $q=1$. Their
physical meaning can only be guessed at but whatever it is, it
should be deep.

\begin{itemize}
\item $q$-Minkowski space has two classical limits, related by
duality. One is the commutative coordinates on $\R^{1,3}$ but
there is another as the homogenized enveloping algebra
$U(su_2\oplus u(1))$.
\item $q$-Minkowski space `quantises' a Poisson-bracket on
$\R^{1,3}$ given by the action of the special conformal
translations.
\item When $q\ne 1$ this action of special conformal
transformations is the braided group adjoint action of
$q$-Minkowski space on itself as an additive braided group.
\end{itemize}

This first item is a version of the general result that the
braided group versions (by transmutation) of the enveloping
algebras $U_q(\cg)$ and their $q$-coordinate algebras are
isomorphic. That is, there is essentially only one object in
$q$-geometry with different scaling limits as $q\to 1$ to give
{\em either} the classical enveloping algebra of $\cg$ {\em or}
the coordinate algebra of $G$. These self-duality isomorphisms
involve dividing by $q-1$ and are therefore singular when $q=1$,
i.e. this is totally alien to conventional geometric ideas. A
homogenized enveloping algebra just means with relations
$\xi\eta-\eta\xi=C[\xi,\eta]$ where the right hand side is the Lie
bracket and $C$ is a central element. In the $q$-deformed case
there is a similar $C=ad-q^2cb$ which is the $q$-Minkowski length;
the mass-shell hyperboloid of $q$-Minkowski space is essentially
the same algebra as $U_q(su_2)$,
\eqn{mshelllim}{\C\left[{\rm mass\ shell}\right]{ {\scriptstyle
1\leftarrow q}\atop
{\ \from \ \atop \ }}\left(\R_q^{1,3}/_{C=1}\right)\isom
U_q(su_2){ {\scriptstyle q\to 1}\atop{ \ \longrightarrow \ \atop\
}}U(su_2)}
\eqn{mshellisom}{\begin{pmatrix}a & b\\ c& d\end{pmatrix}\isom
\begin{pmatrix}q^H& q^{-\h}(q-q^{-1})q^{H\over 2}X_-\\
q^{-\h}(q-q^{-1})X_+q^{H\over 2}&
q^{-H}+q^{-1}(q-q^{-1})^2X_+X_-\end{pmatrix} } which is a
$q$-geometrical point of view on previously known `matrix
generators'\cite{FRT:lie} for quantum groups such as $U_q(su_2)$
-- in our case it comes $q$-geometrically from a covering
isomorphism of $\R_q^{1,3}$ with the braided enveloping algebra
$U(gl_{q,2})$ of a braided-Lie algebra $gl_{q,2}$. Axioms and a
general theory of such Lie algebra objects underlying quantum
groups was one of the important technical achievements of braided
groups in the mid 1990s\cite{Ma:lie}. Notice also that the two
different scaling limits $q\to 1$ exactly implement the Fourier
duality between noncommutative position and constrained momentum
discussed in Section~II.A.

Finally, we note that probably the greatest significance in the
long term of the braided approach is that it solves the following
`uniformity of quantisation problem',
\begin{itemize}
\item There is only one universe.
How do we know when we have quantised this or that space
separately that they are consistent and fit together to a single
quantum universe?
\end{itemize}
The braided approach solves this because we do not start with the
Poisson brackets -- they come from classicalisation -- but from
the deeper principle of braid statistics. Apart from giving the
$q$-deformation of most structures in physics, it does it
uniformly and in a generally consistent way because what what we
deform is actually the category of vector spaces into a braided
category. All constructions based on linear maps then deform
coherently and consistently with each other as braid diagram
constructions (so long as they do not get tangled). After that one
inserts the formulae for specific braidings (e.g. generated by
specific quantum groups) to get the $q$-deformation formulae. {\em
After that} one semiclassicalises by taking commutators to lowest
order, to get the Poisson-bracket that we have just quantised.
Moreover, different quantum groups $U_q(\cg)$ are all mutually
consistent being related to each other by the inductive
construction\cite{Ma:dbos} mentioned at the end of Section~III.A. We
have seen this with the $q$-Lorentz and $q$-conformal groups for
$q$-Minkowski space above.

\section{Quantum manifolds}

In this section we take a closer look at the progress towards the
more general noncommutative differential geometry of which quanutm
groups and braided groups should be a part. If one means by
differential geometry `bundles', `connections', gauge theory etc
then such a theory did emerged in the early 1990s in the work of
T. Brzezinski and the author, with the by-now standard example of
the $q$-monopole. Leading from this there is today a more or less
complete theory that includes most of the naturally occurring
examples but is a general theory not limited to special examples
and models, i.e. has the same degree of `flabbiness' as
conventional geometry.

There {\em are} open problems so it should not be thought that
this section represents the last word; the subject is still
evolving but there is now something on the table. Among other
things, our constructions are purely algebraic with operator and
$C^*$-algebra considerations not fully worked. On the other hand,
there is plenty of concrete motivation. As well as what has
already been said, let us note that as a bonus this programme of
noncommutative geometry will include {\em discrete} geometry as a
special case. It gives a systematic way to do geometry on
lattices, for example, somewhat different from existing ad hoc
methods. This is depicted in Figure~8. Thus, the idea is to find a
general algebraic notion of geometry that includes usual
commutative coordinate algebras as a special case {\em and} that
includes the kind of natural $q$-deformation and other examples
coming out of quantum groups. The latter are a good testing ground
because they have a parameter which we can set $q\to 1$ to verify
the correct classical limit, i.e. we maintain `eye-contact' with
conventional geometry (this is not true of more abstract
approaches based on $C^*$-algebras, for example). Next, when we
are satisfied that we have the natural definitions we can
specialise to finite-dimensional algebras. For example even
commutative ones, which would be differential geometry on finite
sets.
\begin{figure}
\[ \epsfbox{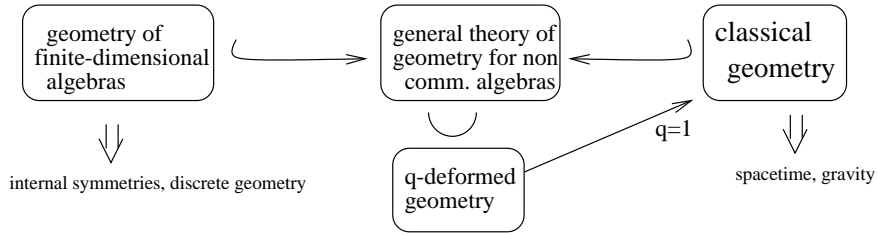}\]
\caption{Noncommutative geometry found with the help of
$q$-deformations can then specialise to finite-dimensional
algebras}
\end{figure}

\begin{itemize}
\item If one can do all of geometry on finite sets then functional
integrals etc. become finite-dimensional integrals etc. and we can
analytically compute the vacuum of QCD, quantum gravity etc. in
the discrete model.
\item We can apply the formalism to finite (commutative or
noncommutative) algebras to provide natural
Dirac operators on quaternions $\quat$ etc. which can figure in
the internal structure of Lagrangians.
\end{itemize}

For example, Connes and J.~Lott at the start of the 1990s proposed
a way to noncommutative-geometrically package the contents of the
standard model in a kind of Kaluza-Klein manner by working with
the coordinate algebra of usual spacetime tensored with a
finite-dimensional algebra $\C\oplus\quat$. In other words,
instead of a compact internal symmetry group one has a
noncommutative space. One can build a `Dirac' operator on the
tensor product from one on the usual spacetime and one on the
finite-dimensional algebra. The latter encodes the mass matrix and
related phenomenological aspects of the theory. This was a novel
approach which gave, in particular, a nice way to think about the
Higgs field mexican-hat potential. On the other hand, from an
abstract point of view almost any matrix can be taken as the
`Dirac' operator on the finite-dimensional algebra so one does not
really get theoretical predictions for the details of that. The
latter can only come from knowing more of the infrastructure of
noncommutative differential geometry to determine what is the
geometrically natural operator to take as Dirac on $\C\oplus
\quat$. While particular applications like this remain to be
developed in the near future, it is clear that for either of the
above reasons the general noncommutative differential geometry,
even specialised to finite-dimensional algebras, can translate
very directly into physical predictions.

\subsection{Quantum differential forms}

We begin here with the first of the different `layers' of
differential geometry, namely the notion of differential structure
itself. Since we omit $C^*$ algebra considerations our coordinate
algebra can be practically any (possibly noncommutative) algebra
$M$. To specify the differential structure we in effect {\em
choose} the cotangent space or differential 1-forms $\Omega^1$.
Since one can multiply forms by `functions' from the left and
right, this should be an $M$-bimodule. There should also be a
linear map $\extd:M\to
\Omega^1$ such that
\eqn{Omega1}{ \extd(ab)=(\extd a)b+a\extd b,\quad \forall a,b\in M}
and $\Omega^1$ should be spanned by elements of the form $a\extd
b$. The main difference from what one might naively take here is
that we do not assume that the left and right multiplications by
$M$ coincide. For if $a\extd b=(\extd b)a$ for all $a,b$, we would
have $\extd[a,b]=0$, which would not be at all suitable for a
generic noncommutative algebra. Differential structures are not
unique even classically, and even more non-unique in the quantum
case. There is, however, one universal example of which others are
quotients. This is
\eqn{univ}{ \Omega^1_{\rm univ}=\ker\cdot\subset M\tens M,\quad \extd
a=a\tens 1-1\tens a.} The universal calculus was studied by
algebraic topologists in the 1970s and is common to practically
all approaches to noncommutative geometry. Our first task is to
choose an appropriate quotient.

Classically, we do not think about this much because on a group
there {\em is} a unique translation-invariant differential
calculus; since we generally work with manifolds built on or
closely related to groups we tend to take the inherited
differential structure without thinking. In the quantum case, i.e.
when $M$ is a quantum group one has a similar notion: a
differential calculus is bicovariant if there are coactions
$\Omega^1\to \Omega^1\tens M, \Omega^1\to M\tens \Omega^1$
forming a bicomodule and compatible with the bimodule structures
and $\extd$. This natural definition was proposed by
Woronowicz\cite{Wor:dif} at the end of the 1980s along with a
couple of examples, but it took another decade before systematic
classification results for the possible such calculi appeared
\cite{Ma:cla}. By now the complete range of possibilities for all
main classes of Hopf algebras are more or less understood. We
begin with a sample of the general results, taken from
\cite{Ma:fie}\cite{Ma:cla} by the author.

For the one-dimensional case of polynomials $M=k[x]$ with values
in any field $k$, the coirreducible calculi (those with no
further quotients) have the form $\Omega^1=k_\lambda[x]$ where
$k_\lambda$ is a field extension of the form $k[\lambda]$ modulo
$m(\lambda)=0$ and $m$ is an irreducible monic polynomial. The
differential and bimodule structures are \eqn{diffline}{  \extd
f(x)=\frac{f(x+\lambda)-f(x)}{\lambda},\quad f(x)\cdot
g(\lambda,x)=f(x+\lambda)g(\lambda,x),\quad g(\lambda,x)\cdot
f(x)=g(\lambda,x)f(x)} for functions $f$ and one-forms $g$. For
example, the calculi on $\C[x]$ are classified by $\lambda_0\in
\C$ (here $m(\lambda)=\lambda-\lambda_0$) and one has
\eqn{diffC}{ \Omega^1=\extd x\C[x],\quad \extd f=\extd
x\frac{f(x+\lambda_0)-f(x)}{\lambda_0},\quad x\extd x=(\extd
x)x+\lambda_0.} We see that the Newtonian case $\lambda_0=0$ is
only one special point in the moduli space of quantum differential
calculi. But if Newton had not supposed that differentials and
forms commute he would have had no need to take this limit. What
one finds with noncommutative geometry is that there is no need to
take this limit at all. In particular, noncommutative geometry
extends our usual concepts of geometry to lattice theory without
taking the limit of the lattice spacing going to zero. It is also
interesting that the most important field extension in physics,
$\R\subset\C$, can be viewed noncommutative-geometrically with
complex functions $\C[x]$ the quantum 1-forms on the algebra of
real functions $\R[x]$.

We can similarly consider functions $M=\C[G]$ on a finite group.
Then the coirreducible calculi correspond to nontrivial conjugacy
classes $\CC\subset G$ and have the form
\eqn{diffC(G)}{ \Omega^1=\CC\cdot\C[G],\quad \extd f
=\sum_{g\in\CC}g\cdot (L_g(f)-f),\quad f\cdot g=g\cdot L_g(f)}
where $L_g(f)=f(g\cdot)$ is the translate of $f$. The cases of
$\C[x]$ and $\C[G]$ are trivial enough to have been observed by
hand before the general classification theorems arrived.

Next, for true examples of noncommutative coordinates we can
consider $M=\C G$ generated by a finite group. It is the dual of
the $\C[G]$ case but regarded `up side down' as a noncommutative
space when $G$ is nonAbelian. One has that the coirreducible
calculi correspond to pairs $(V,\rho,\lambda)$ where $(V,\rho)$
is a nontrivial irreducible representation and $\lambda\in V/\C$.
They have the form \eqn{diffCG}{ \Omega^1=V\cdot \C G,\quad
\extd  g =((\rho(g)-1)\lambda)\cdot g,\quad g\cdot
v=(\rho(g)v)\cdot g} where $g\in G$ is regarded as a `function'.
For $M=U(\cg)$ as in Section~II one has a similar construction
for any irreducible representation $V$ of the Lie algebra $\cg$
and choice of ray $\lambda$ in it. The complete classification in
the Lie cases of either the enveloping or the coordinate algebra
are not known.

This covers the classical objects or their duals. For a
finite-group bicrossproduct $\C[M]\bicross \C G$ the
classification is a mixture of the two cases above and is given in
\cite{BegMa:dif}. Another method covers the Planck scale quantum
group $\C[x]\bicross\C[p]$ in Section~II, namely this is a
twisting by a cocycle of its classical limit $\C[\R\rcross\R]$
and there is a theorem that its differentials are thereby in
correspondence with those in the classical limit\cite{MaOec:twi}.
Finally, we come to the quantum groups like $\C_q[G]$ as in
Section~III. Here there is a theorem\cite{Ma:cla} that the
coirreducible calculi are basically in correspondence with
nontrivial irreducible representations $V$ of the Lie algebra and
have the form \eqn{diffqG}{ \Omega^1=(V\tens V^*)\cdot\C_q[G].}
In fact one has one natural calculus for each irreducible
representation plus some `shadows' or technical variants allowed
according to the precise formulation of the relevant quantum
groups and their duality (this is more a deficit in the technical
definitions than anything else). For example, for $\C_q[SU_2]$
there is basically one bicovariant calculus for each spin $j$
with dimension $(2j+1)^2$. The lowest is 4 dimensional. Actually
the spanning vector space here is the 4-dimensional braided-Lie
algebra $gl_{q,2}$ mentioned at the end of Section~III.D. It is
irreducible for generic $q$ but as $q\to 1$ it degenerates into
$su_2\oplus u(1)$. Correspondingly the 4-dimensional calculus
becomes a direct sum of the usual 3-dimensional calculus on $SU_2$
and an additional operator, the Casimir operator or Laplacian:
\begin{itemize}
\item Bicovariant $\Omega^1$ for all main classes of quantum
groups have been classified
\item In particular, when we $q$-deform $SU_2$ its usual
differentials and its Laplacian are necessarily bound up in one
coirreducible 4-dimensional bicovariant calculus.
\end{itemize}
Other non-bicovariant calculi are possible also, including a
standard 3-dimensional left-covariant calculus on $\C_q[SU_2]$
known since the original work of Woronowicz.

Finally, on a quantum group there is a natural extension to higher
order forms and in fact an entire exterior algebra once the
bicovariant 1-forms have been chosen\cite{Wor:dif}. Other
extensions are also possible. Given the extension, one has a
quantum cohomology defined in the usual way as closed forms modulo
exact ones. To close with one non-quantum group example, consider
any actual manifold with a finite good cover $\{U_i\}_{i\in I}$.
Instead of building geometric invariants on a manifold and
studying them modulo diffeomorphisms we can use the methods above
to first pass to the skeleton of the manifold defined by its open
set structure and do differential geometry directly on this
indexing set $I$. Thus we take $M=\C[I]$ which just means
collections $\{f_i\in \C\}$. The universal $\Omega^1$ is just
matrices $\{f_{ij}\}$ vanishing on the diagonal. We use the
intersection data for the open sets to set some of these to zero.
Similarly for higher forms. Thus\cite{BrzMa:dif}
\[\Omega^1=\{f_{ij}|\ U_i\cap U_j\ne \emptyset\},\quad
\Omega^2=\{f_{ijk}|\ U_i\cap U_j\cap U_k\ne\emptyset\}\]
\eqn{cech}{(\extd f)_{ij}=f_i-f_j,\quad (\extd f)_{ijk}
=f_{ij}-f_{ik}+f_{jk}}
and so on. Then one has that the quantum cohomology is just the
additive Cech cohomology of the original manifold.

Apart from cohomology one can start to do gauge theory, at least
with trivial bundles. At this level a `U(1)' gauge field is just a
differential form $\alpha\in \Omega^1$ and its curvature is
$F=\extd\alpha+\alpha\wedge\alpha$, etc. A gauge transform is
\eqn{trivialgauge}{ \alpha^\gamma=\gamma^{-1}\alpha\gamma
+\gamma^{-1}\extd \gamma,\quad F^\gamma=\gamma^{-1} F\gamma}
for any invertible `function' $\gamma\in M$, and so on. One can
certainly obtain interesting results even when the base is
classical (but the calculus is quantum). For the example
associated to open sets one has that the zero curvature gauge
fields modulo gauge transformations recovers again the first Cech
cohomology, but now in a multiplicative form. Or by chosing a
quantum calculus even on usual $\R^n$ it is clear that nonlinear
and higher-derivative equations could be viewed as zero curvature
ones. Solutions would typically then be provided by
$\alpha=\gamma^{-1}\extd\gamma$ i.e., pure gauge. There have been
some first efforts in this direction in the physics literature. It
should be clear at least that noncommutative differentials have
the potential to unify and make clearer a whole range of otherwise
ad-hoc constructions ranging from group theory to number theory to
lattice differentials and integrable systems.

\subsection{Bundles and connections}

The next layer of differential geometry is bundles, connections,
etc. Usually in physics one needs only the local picture with
trivial bundles in each open set -- but for a general
noncommutative algebra $M$ there may be no reasonable `open sets'
and one has therefore to develop the global picture from the
start. We need nontrivial bundles to cover physics such as in the
Bohm-Aharanov effect, potential effects such as the monopole and
also to cover homogeneous spaces and the frame bundles of general
`manifolds'. None of these could be understood without a global
point of view. In particular, the next quantum spaces after
quantum groups and quantum-braided planes are quanutm homogeneous
spaces and examples such as the quantum sphere $\C_q[S^2]$
(actually a two-parameter family of them) were known already by
the end of the 1980s \cite{Pod:sph}. The required noncommutative
differential geometry to really understand them as bundles came a
few years later in \cite{BrzMa:gau}.

Also note that this is a different problem from going from $U(1)$
gauge theory to nonAbelian, but the two can be handled together,
i.e. we want a general quantum group as gauge group. For trivial
bundles we could just take a gauge field as $\alpha\in
\Omega^1\tens U_q(\cg)$ for example and write down similar
formulae to those at the end of the last section. A gauge
transform is an invertible element $\gamma\in M\tens U_q(\cg)$
etc. For our global geometrical picture however, we need to think
of the quantum group geometrically and work with $H$ viewed as
more like $\C[G]$ or a $q$-coordinate algebra $\C_q[G]$ etc. This
is the setting for the present section. Also, to keep things
simple we give formulae only for the universal differential
calculus but the general case is also covered by making suitable
quotients.

Basically, a classical bundle has a free action of a group and a
local triviality property. In our algebraic terms we need an
algebra $P$ in the role of `coordinate algebra of the total space
of the bundle' and a coaction $\Delta_R:P\to P\tens H$ of the
quantum group $H$ such that the fixed subalgebra is $M$,
\eqn{fixed}{ M=P^H=\{p\in P|\ \Delta_R p=p\tens 1\}.}
Freeness and local triviality are replaced by the requirement that
\eqn{exactness}{ 0\to P(\Omega^1M)P\to\Omega^1P{\buildrel{\rm ver}
\over\longrightarrow} P\tens \ker\eps\to 0}
is exact, where ${\rm ver}=(\cdot\tens\id)\Delta_R$ plays the role
of generator of the vertical vector fields corresponding
classically to the action of the group (for each element of $H^*$
it maps $\Omega^1P\to P$ like a vector field). Exactness on the
left says that the one-forms $P(\Omega^1M)P$ lifted from the base
are exactly the ones annihilated by the vertical vector fields.

One can then define a connection as an equivariant splitting
\eqn{connection}{ \Omega^1P=P(\Omega^1 M)P\oplus {\rm complement}}
i.e. an equivariant projection $\Pi$ on $\Omega^1P$. One can show
the required analogue of the usual theory, i.e. that such a
projection corresponds to a connection form
\eqn{conform}{\omega:\ker\eps\to\Omega^1P,\quad {\rm ver}\circ
\omega=1\tens\id} where $\omega$ intertwines with the adjoint
coaction of $H$ on itself. Finally, one can define associated
bundles. If $V$ is a vector space on which $H$ coacts then we
define the associated `bundles' $E^*=(P\tens V)^H$ and
$E=\hom_H(V,P)$, the space of intertwiners. The two bundles
should be viewed geometrically as `sections' in classical
geometry of bundles associated to $V$ and $V^*$. Given a suitable
(strong) connection one has a covariant derivative
\eqn{covderiv}{ D_\omega:E\to E\tens_M \Omega^1M,\quad D_\omega
=(\id-\Pi)\circ \extd}

All of this can be checked out on the $q$-monopole bundle over the
$q$-sphere\cite{BrzMa:gau}. Recall that classically the inclusion
$U(1)\subset SU_2$ in the diagonal has coset space $S^2$ and
defines the $U(1)$ bundle over the sphere on which the monopole
lives. The same idea works here, but since we deal with coordinate
algebras the arrows are reversed. The coordinate algebra of $U(1)$
is the polynomials $\C[g,g^{-1}]$ and the classical inclusion
becomes the projection
\eqn{sphproj}{\pi:\C_q[SU_2]\to \C[g,g^{-1}],\quad
\pi\left(\begin{matrix}
a&b\\ c&d
\end{matrix}\right)=\left(\begin{matrix}
g&0\\ 0&g^{-1}
\end{matrix}\right)}
Its induced coaction $\Delta_R=(\id\tens\pi)\Delta$ is by the
degree defined as the number of $a,c$ minus the number of $b,d$ in
an expression. The quantum sphere $\C_q[S^2]$ is the fixed
subalgebra i.e. the degree zero part. Explicitly, it is generated
by $b_3=ad$, $b_+=cd$, $b_-=ab$ with $q$-commutativity relations
\eqn{qsph}{ b_\pm b_3=q^{\pm 2}b_3b_\pm+(1-q^{\pm 2})b_\pm, \quad
q^{2}b_-b_+=q^{-2}b_+b_-+(q-q^{-1})(b_3-1)} and the sphere
equation $b_3^2=b_3+qb_-b_+$. When $q\to 1$ we can write
$b_\pm=\pm(x\pm\imath y)$, $b_3=z+\h$ and the sphere equation
becomes $x^2+y^2+z^2=\frac{1}{4}$ while the others become that
$x,y,z$ commute. It turns out that we have a quantum bundle in
the sense above and that there is a connection
$\omega(g-1)=d\extd a-qb\extd c$ which, as $q\to 1$, becomes the
usual Dirac monopole constructed algebraically.

It is easy to see that the `matter fields' or sections of the
associated vector bundles $E_n$ for each charge $n$ are just the
degree $n$ parts of $\C_q[SU_2]$. The associated covariant
derivative acts on these. This is also where the noncommutative
differential geometry coming out of quantum groups links up with
the more traditional $C^*$-algebra approach of A. Connes and
others. Traditionally a vector bundle over any algebra is defined
as a finitely generated projective module. However, there was no
notion of quantum principal bundle before quantum groups. The
associated bundles $E_n$ for the $q$-monopole bundle indeed turned
out to be finitely generated projective modules\cite{HajMa:pro},
i.e. there is an $(|n|+1)\times(|n|+1)$-matrix $e_n$ with values
in $\C_q[S^2]$ with $e_n^2=e_n$ and $E_n=e_n\C_q[S^2]^{|n|+1}$.
The covariant derivative for the monopole in these terms has the
form $e_n\extd e_n$. For the lowest charge the projector is
\eqn{projmon}{e_1=\begin{pmatrix} b_3&-qb_-\\ b_+& q^2(1-b_3)
\end{pmatrix}.}
The projectors are elements of the noncommutative $K$-theory
$K_0(\C_q[S^2])$ and have a duality pairing with Connes' cyclic
cohomology\cite{Con:geo} which for the $q$-monopole gives the
correct answer as its Chern class. Thus the quantum groups
approach ties up in the end with Connes' approach but provides
more of the (so far algebraic) infrastructure of differential
geometry
-- principal bundles, connection forms, etc. otherwise missing.

The potential applications of quantum group gauge theory hardly
need to be elaborated. Among the more esoteric let us note that
nonAbelian gauge fields provide invariants of manifolds and hence
similarly one could obtain `geometric' invariants of
noncommutative algebras $M$. For example, for a classical manifold
\eqn{pi1}{ \left\{ {{{\rm Flat\ connections\ on}\ G-{\rm bundle}}
\atop{\rm modulo\ gauge}}\right\}\isom \hom(\pi_1,G)/G}
using the holonomy. One can view this as a functor from groups to
sets and the homotopy group $\pi_1$ as more or less the
representing object in the category of groups. The same idea with
quantum group gauge theory defines $\pi_1(M)$ as a homotopy
quantum group for any algebra $M$ as more or less the representing
object of the functor that assigns to a quantum group $H$ the set
of zero-curvature gauge fields with this quantum structure group.
This goes somewhat beyond vector bundles and $K$-theory alone.
Although in principle defined, this idea has yet to be developed
in a computable form. It is one of many directions for the future.
Other directions include discrete e.g. finite models of QCD and
functional integration in this setting.

Finally we mention that one needs to make a slight generalisation
of the above to include other noncommutative examples of interest.
In fact (and a little unexpectedly) the general theory above can
be developed with only a coalgebra rather than a Hopf algebra $H$.
Or dually it means only an algebra $A$ in place of the enveloping
algebra of a Lie algebra. This was achieved more recently, in
\cite{BrzMa:coa}\cite{BrzMa:geo}, and allows us to include the
full 2-parameter quantum spheres as well as (in principle) to all
known $q$-deformed symmetric spaces. Beyond that, we can as
mentioned apply the theory to our favourite finite-dimensional
algebras or to commutative algebras with quantum differential
calculi, etc. Again it seems likely that some startling
applications along these lines will emerge in coming years.

\subsection{Quantum soldering forms and metrics}

We are finally ready to take the plunge and define a `quantum
manifold'. If our primary goal is to unify quantum theory and
gravity through some noncommutative generalisation of geometry
then the following at least puts something on the table to try
out. This theory was only recently proposed in \cite{Ma:rie} and
has therefore been little explored so far. But it does already
predict a slight generalisation even of conventional Riemannian
geometry as naturally appearing by semiclassicalisation. One could
use conventional geometric methods to first explore the classical
predictions of that generalisation even before getting into the
noncommutative theory. The approach we take is basically that of a
vierbein or, in global terms, a soldering form. This expresses
gravity as a gauge theory of the frame bundle so that we can use
the formalism of the previous section.

The first step is to define a generalised frame bundle or {\em
frame resolution} of our algebra $M$ as a quantum principal bundle
$(P,H,\Delta_R)$ over $M$, a comodule $V$ and an equivariant
`soldering form' $\theta:V\to P\Omega^1M\subset\Omega^1P$ such
that the induced map
\eqn{frame}{ E^*\to \Omega^1M,\quad p\tens v\mapsto p\theta(v)}
is an isomorphism. What this does is to express the cotangent
bundle as associated to a principal one. Other tensors are then
similarly associated, for example vector fields are
$E\isom\Omega^{-1}M$. Of course, all of this has to be done with
suitable choices of differential calculi on $M,P,H$ whereas we
have been focusing for simplicity on the universal calculi. There
are some technicalities here but more or less the same definitions
work in general. The working definition\cite{Ma:rie} of a {\em
quantum manifold} is simply this data
$(M,\Omega^1,P,H,\Delta_R,V,\theta)$. The definition works in that
one has analogues of many usual results. For example, a connection
$\omega$ on the frame bundle induces a covariant derivative
$D_\omega$ on the associated bundle $E^*$ which maps over under
the soldering isomorphism to a covariant derivative
\eqn{nabla}{ \nabla:\Omega^1M\to \Omega^1M\tens_M\Omega^1M.}
Its torsion is defined as corresponding similarly to
$D_\omega\theta$.

Defining a Riemannian structure is harder. It turns out that it
can be done in a `self-dual' manner as follows. Given a framing, a
`generalised metric' isomorphism $\Omega^{-1}M\isom\Omega^1M$
between vector fields and one forms can be viewed as the existence
of {\em another} framing $\theta^*:V^*\to(\Omega^1M)P$, which we
call the {\em coframing}, this time with $V^*$. Nondegeneracy of
the metric corresponds to $\theta^*$ inducing an isomorphism
$E\isom\Omega^1M$. The working definition of a quantum Riemannian
manifold is therefore the data $(M,\Omega^1,
P,H,\Delta_R,V,\theta,\theta^*)$, where we have a framing and at
the same time $(M,\Omega^1,P,H,\Delta_R,V^*,\theta^*)$ is another
framing. The associated quantum metric is
\eqn{quametric}{ g= \theta^*(f^a)\theta(e_a)\in\Omega^1M
\tens_M\Omega^1M}
where $\{e_a\}$ is a basis of $V$ and $\{f^a\}$ is a dual basis
(c.f. the canonical element $\exp$ from Fourier theory in
Section~II.C).

Now, this self-dual formulation of `metric' as framing and
coframing is symmetric between the two. One could regard the
coframing as the framing and vice versa. From our original point
of view its torsion tensor corresponding to $D_\omega\theta^*$ is
some other tensor, which we call the {\em cotorsion tensor}. This
is a new concept which did not exist in conventional differential
geometry. We then define a generalised Levi-Civita connection on a
quantum Riemannian manifold as the $\nabla$ of a connection
$\omega$ such that the torsion and cotorsion tensors both vanish.
The Riemannian curvature of course corresponds to the curvature of
$\omega$, which is $\extd\omega+\omega\wedge\omega$, via the
soldering form. I would not say that the Ricci tensor and Einstein
tensor are understood abstractly enough in this formalism but of
course one can just write down the relevant contractions and
proceed blindly.

This is about as far as the programme has come at present. It is
known\cite{Ma:rie} that every quantum group with bicovariant
calculus is a quantum manifold in this sense. And for quantum
groups such as $\C_q[SU_2]$ there is an Ad-invariant
non-degenerate braided Killing form on the braided-Lie algebra
$gl_{q,2}$ in \cite{Ma:lie} which provides a coframing from a
framing -- so that quantum groups such as $\C_q[SU_2]$ with such
differential calculi are quantum Riemannian manifolds in the
required sense. At least with the universal calculus every quantum
homogeneous space is a quantum manifold too. That includes quantum
spheres, quantum planes etc. In fact, there is a notion of
comeasuring or automorphism quantum group\cite{Ma:dif} for
practically any algebra $M$ and when this has an antipode (which
typically requires some form of completion) one can write $M$ as
a quantum homogeneous space. So any $M$ is more or less a quantum
manifold for some principal bundle (at least rather formally).
This is analogous to the idea that any manifold is, rather
formally, a homogeneous space of diffeomorphisms modulo
diffeomorphisms fixing a base point. So the formalism does appear
to be `flabby' or general enough to sensibly write down field
equations etc.

Finally, to get the physical meaning of the cotorsion tensor and
other ideas coming out of this noncommutative Riemannian geometry,
let us consider the semiclassical limit. What we find is that
noncommutative geometry forces us to slightly generalise
conventional Riemannian geometry itself. If noncommutative
geometry is closer to what comes out of quantum gravity then this
generalisation of conventional Riemannian geometry should be
needed to include Planck scale effects or at least to be
consistent with them when they emerge at the next order of
approximation. The generalisation, more or less forced by the
noncommutativity, is as follows:

\begin{itemize}
\item We have to allow any group $G$ in the `frame bundle', hence
the more general concept of a `frame resolution'
$(P,G,V,\theta_\mu^a)$ or {\em generalised manifold}.

\item The {\em generalised metric} $g_{\mu\nu}
=\theta^*_\mu{}^a\theta_{\nu a}$
corresponding to a coframing $\theta^*_{\mu}{}^a$ is nondegenerate
but need not be symmetric.

\item The {\em generalised Levi-Civita} connection defined as having
vanishing torsion and vanishing cotorsion respects the metric only
in a skew sense
\eqn{genlev}{ \nabla_\mu g_{\nu\rho}-\nabla_\nu g_{\mu\rho}=0}

\item The group $G$ is not unique (different flavours of frames are
possible, e.g. an $E_6$-resolved manifold), not necessarily based
on $SO_n$. This gives different flavours of covariant derivative
$\nabla$ that can be induced by a connection form $\omega$.

\item Even when $G$ is fixed and $g_{\mu\nu}$ is fixed, the
generalised Levi-Civita condition does not fix $\nabla$ uniquely,
i.e. one should use a first order $(g_{\mu\nu},\nabla)$ or
$(\theta,\theta^*,\omega)$ formalism.
\end{itemize}

To explain (\ref{genlev}) we should note the general result
\cite{Ma:rie} that for any generalised metric one has
\eqn{torcotor}{\nabla_\mu g_{\nu\rho}-\nabla_\nu g_{\mu\rho}={\rm
CoTorsion}_{\mu\nu\rho}-{\rm Torsion}_{\mu\nu\rho},} where we use
the metric to lower all indices.

This generalisation of Riemannian geometry includes special cases
of symplectic geometry, where the generalised metric is totally
antisymmetric. So the two are unified in our formulation, which is
what we would expect if the theory is to be the
semiclassicalisation of a theory unifying quantum theory and
geometry. It is also remarkable that metrics with antisymmetric
part are exactly what are needed in string theory to establish
T-duality. In summary, one has on the table a general
noncommutative Riemannian geometry to play with. One can try it
out on simple examples such as quaternions or on discrete spaces
(for example doing a functional integral as a finite-dimensional
integral to do quantum gravity). This is a direction for ongoing
work at present. One can also explore the duality ideas of
Section~II. In particular the observable-state duality should
translate into a relation between gravity and thermodynamical
entropy in this algebraic setting. Finally, it is at least clear
that there are immediate predictions even at the classical level
in terms of a classical generalisation of Riemannian geometry with
antisymmetric parts to the metric, something with can be explored
using conventional geometric methods. This is more or less the
state of the art at the time of writing. While I doubt that these
are anything more than first exploratory efforts, it does seem
that something {\em like} this has to be a step in the right
direction for the unification of quantum theory and gravity.

\section{A new philosophical foundation for the next millennium}

Since the millennium happens only once in a thousand years, this
may now be a good time to sit back and think a bit about the long
long term implications (if any) of what we are doing. After all,
lay persons turn to physicists for insights into the nature of
physical reality. Apart from technical calculations, do we have
anything to really tell them? While I doubt that quantum groups
and noncommutative geometry are the end of the line -- i.e. even
more powerful concepts will be needed later on, they do
demonstrate some general ideas which I will try to explain here.
These more general ideas are, in a nutshell, about the nature of
the relationship between mathematics and physics. It seems clear
to me that, on top of technical advancements, a future revolution
in our understanding of Nature will probably also need new
philosophical input and so we should not shy away from thinking
about that. I have saved this best part for the last, not least
because it is of necessity very much my personal view even more
than previous sections. Most of it was eventually published in my
1987 essay on the nature of physical reality\cite{Ma:pri} and it
may be considered as background motivation for almost all of my
own work since then.

A safe starting point should be that whatever we may say today
about fundamental physics that is based on our past experience and
not on the deepest philosophical principles is not likely to be
correct or to survive very far into the next millennium other than
as an approximation. So as a basis we should stick only to some of
the deepest principles. In my opinion one of the deepest
principles concerns the nature of mathematics itself. Namely,
throughout mathematics one finds an intrinsic dualism between
observer and observed as follows. When we think of a function $f$
being evaluated on $x\in X$, we could equally-well think of the
same numbers as $x$ being evaluated on $f$ a member of some dual
structure $f\in\hat X$:
\eqn{gelf}{ {\rm Result}=f(x)=x(f).}
Such a `turning of the tables' is a mathematical fact. For any
mathematical concept $X$ one may consider maps or
`representations' from it to some self-evident class of objects
(say rational numbers or for convenience real or complex numbers)
wherein our results of measurements are deemed to lie. Such
representations themselves form a dual structure $\hat X$ of which
elements of $X$ can be equally well viewed as representations.
{\em But is such a dual structure equally real?} It was postulated
in \cite{Ma:pri} that indeed this should be so in a complete
theory,
\begin{itemize}
\item The search for a complete theory of physics is the search
for a self-dual formulation in the above representation-theoretic
sense (the principle of representation-theoretic self-duality)
\end{itemize}
Put another way, a complete theory of physics should admit a
`polarisation' into two halves each of which is the set of
representations of the other. This division should be arbitrary --
one should be able to reverse interpretations (or indeed consider
canonical transformations to other choices of `polarisation' if
one takes the symplectic analogy).

Note that by completeness here I do not mean knowing in more and
more detail {\em what} is true in the real world. That consists of
greater and greater complexity but it is not {\em theoretical}
physics. I'm considering that a theorist wants to know {\em why}
things are the way they are. Ideally I would like on my deathbed
to be able to say that I have found the right point of view or
theoretical-conceptual framework from which everything else
follows. Working out the details of that would be far from trivial
of course. We are taking at this point a more or less conventional
reductionist viewpoint except that the Principle asserts that we
will not have found the required point of view unless it is
self-dual in the above sense.

We have already seen in Section~II how such a general philosophy
could translate in practice. In the setting of quantum gravity we
take the view first of all that geometry -- or `gravity' is dual
to quantum theory or matter. We discussed this for simple models
such as spheres with constant curvature where it was achieved by
Fourier theory. If we accept this then in general terms Planck
scale physics has to unify these mutually dual concepts into one
structure. Einstein's equation \eqn{einst}{ G_{\mu\nu}\propto
T_{\mu\nu}} may then even appear as some kind of self-duality
equation within this self-dual context. Here the stress-energy
tensor $T_{\mu\nu}$ relates to how matter responds to the
geometry, while the Einstein tensor $G_{\mu\nu}$ measures how
geometry responds to matter. This is the part of Mach's principle
which apparently inspired Einstein. To do it properly one needs
clearly some kind of noncommutative geometry because $T_{\mu\nu}$
should really be the quantum operator stress-energy and its
coupling to $G_{\mu\nu}$ through its expectation value is surely
only the first approximation or semiclassical limit of an
operator version of (\ref{einst}). But an operator version of
$G_{\mu\nu}$ only makes sense in the context of noncommutative
geometry. What we would hope to find, in a suitable version of
these ideas, is a self-dual setting where there is a dual
interpretation in which $T_{\mu\nu}$ is the Einstein tensor of
some dual system and $G_{\mu\nu}$ its stress-energy. In this way
the duality and self-duality of the situation would be made
manifest.

While the above remarks cannot yet be made fully precise at that
level of generality, quantum groups provide a simple and soluble
version of this unification problem. \begin{figure}
\[ \epsfbox{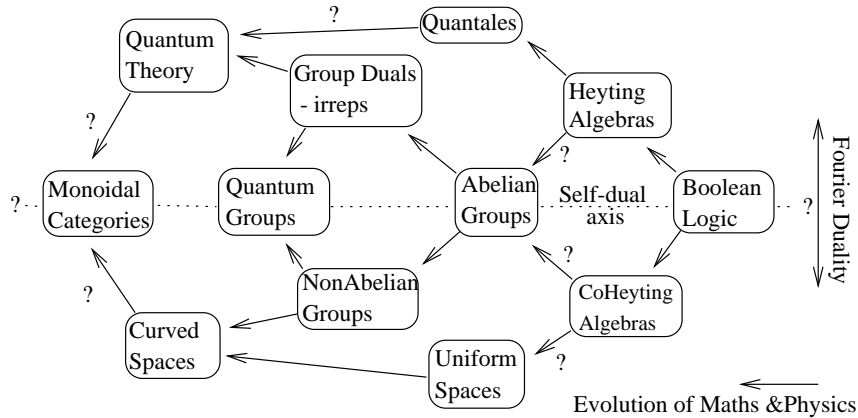}\]
\caption{Representation-theoretic approach to Planck-scale physics.
The unification of quantum and geometrical effects is a drive to
the self-dual axis. Arrows denote inclusion functors}
\end{figure}
This is shown in Figure~9. Thus, the simplest theories of physics
are based on Boolean algebras (a theory consists of classification
of a `universe' set into subsets); there is a well-known duality
operation interchanging a subset and its complement. The next more
advanced self-dual category is that of (locally compact) Abelian
groups such as $\R^n$. In this case the set of 1-dimensional
(ir)reps is again an Abelian group, i.e. the category of such
objects is self-dual. In the topological setting one has
$\hat\R^n\isom \R^n$ so that these groups (which are at the core
of linear algebra) are self-dual objects in the self-dual category
of Abelian groups. Of course, Fourier theory interchanges these
two. More generally, to accommodate other phenomena we step away
from the self-dual axis. Thus, nonAbelian Lie groups such as
$SU_2$ as manifolds provide the simplest examples of curved
spaces. Their duals, which means constructing irreps, appear as
central structures in quantum field theory (as judged by any
course on particle physics in the 1960s). Wigner even defined a
particle as an irrep of the Poincar\'e group. The unification of
these two concepts, groups and groups duals was for many years an
open problem in mathematics. Hopf algebras or quantum groups had
already been invented in the 1940s and provided in particular the
next more general self-dual category containing groups and group
duals (or both coordinate algebras $\C[G]$ and enveloping algebras
$U(\cg)$) in which to attempt this unification. So the language
existed but the problem to find examples of true quantum groups
going genuinely beyond these and unifying them was open. It is
remarkable that at the same time as this mathematical problem was
going on, the problem of unifying quantum theory and gravity was
going on in physics. Moreover, in a self-dual category we can even
go further and look for self-dual objects as a further constraint
on the detailed structure of the model. This was the thinking
behind the bicrossproduct quantum groups in
Section~II.E\cite{Ma:the}\cite{Ma:phy}. We saw that the
Planck-scale Hopf algebra indeed has both quantum and geometrical
features and detailed structure, including dynamics not unlike a
black-hole event horizon, coming out of the self-duality
constraint. We also saw how the self-duality can be understood
physically as an observable-state symmetry as in (\ref{gelf})
between the geometry and the quantum aspects.

On the other hand the kind of general principle listed above is
not just tied to this one setting. It could in principle both help
predict the structure of more advanced theory of physics and, with
hindsight, help us to conceptually organise its already know
structure. This is because the structure of the theory of
self-dual structures is nontrivial and not everything is possible.
Knowing what is mathematically possible and combining with some
postulates such as the above is not empty. For example, back in
1989 and motivated in the above manner it was shown that the
category of monoidal categories (i.e. categories equipped with
tensor products) was itself a self-dual category, i.e. that there
was a construction $\hat{\CC}$ for every such category
$\CC$\cite{Ma:rep}. Since then it has turned out that both
conformal field theory and certain other quantum field theories
can indeed be expressed in such categorical terms. Geometrical
constructions can also be expressed categorically\cite{Ma:som}. On
the other hand, this categorical approach is still under-developed
and its exact use and the exact nature of the required duality as
a unification of quantum theory and gravity is still open. I would
claim only `something like that' (one should not expect too much
from philosophy alone).

Another point to be made from Figure~9 is that if quantum theory
and gravity already take us to very general structures such as
categories themselves for the unifying concept then, in lay terms,
what it means is that the required theory involves very general
concepts indeed of a similar level to semiotics and linguistics
(speaking about categories of categories etc.). It is almost
impossible to conceive {\em within existing mathematics} (since it
is itself founded in categories) what fundamentally more general
structures would come after that. In other words, {\em the
required mathematics is running out} it least in the manner that
it was developed in this century (i.e. categorically) and at least
in terms of the required higher levels of generality in which to
look for self-dual structures. If the search for the ultimate
theory of physics is to be restricted to logic and mathematics
(which is surely what distinguishes science from, say, poetry),
then this indeed correlates with our physical intuition that the
unification of quantum theory and gravity is the last big
unification for physics as we know it, or that theoretical physics
as we know it is coming to an end. I would agree with this
assertion except to say that the new theory will probably open up
more questions which are currently considered metaphysics and make
them physics, so I don't really think we will be out of a job even
as theorists (and there will always be an infinite amount of
`what' work to be done even if the `why' question was answered at
some consensual level).

As well as seeking the `end of physics', we can also ask more
about its birth. Again there are many nontrivial and nonempty
questions raised by the self-duality postulate. Certainly the key
generalisation of Boolean logic to intuitionistic logic is to
relax the axiom that $a\cup\tilde a=1$ (that $a$ or not $a$ is
true). Such an algebra is called a {\em Heyting algebra} and can
be regarded as the birth of quantum mechanics. Dual to this is the
notion of a {\em coHeyting algebra} in which we relax the law that
$a\cap\tilde a=0$. In such an algebra one can define the
`boundary' of a proposition as
\eqn{cohey}{ \del a=a\cap \tilde a}
and show that it behaves like a derivation. This is surely the
birth of geometry. How exactly this complementation duality
extends to the Fourier duality for groups and on to the duality
between more complex geometries and quantum theory is not
completely understood, but there are conceptual `physical'
arguments that this should be so, put forward in \cite{Ma:pri}.
Thus, in the simplest `theories of physics' based only on logic
one can work equally well with `apples' or `not-apples' as the
names of subsets.
\begin{itemize}
\item What happens to this complementation
duality in more advanced theories of physics? Apples curve space
while not-apples do not, i.e. in physics one talks of apples as
really existing while not-apples are merely an abstract concept.
\end{itemize}
Clearly the self-duality is lost in a theory of gravity alone. But
we have argued\cite{Ma:pri} that when one considers both gravity
{\em and} quantum theory, the self-duality can be restored. Thus
when we say that a region is as full of apples as General
Relativity allows (more matter simply forms a black hole which
expands), which is the right hand limiting line in Figure~10, in the dual
theory we might say that the region is as empty of not-apples as
quantum theory allows, the limitation being the left slope in
Figure~10. Here the uncertainty principle in the form of pair
creation ensures that space cannot be totally empty of
`particles'. Although heuristic, these are arguments that quantum
theory and gravity are dual and that this duality is an extension
of complementation duality. Only a theory with both would be
self-dual. Also, in view of a `hole' moving in the opposite
direction to a particle, the dual theory should also involves time
reversal. The self-duality is something like CPT invariance but in
a theory where gravitational and not only quantum effects are
considered. We are proposing it as a key requirement for
quantum-gravity. Diagrams similar to the right hand side of Figure~10
have been attributed to Brandon Carter as a tool to plot stellar evolution.

\begin{figure}
\[ \epsfbox{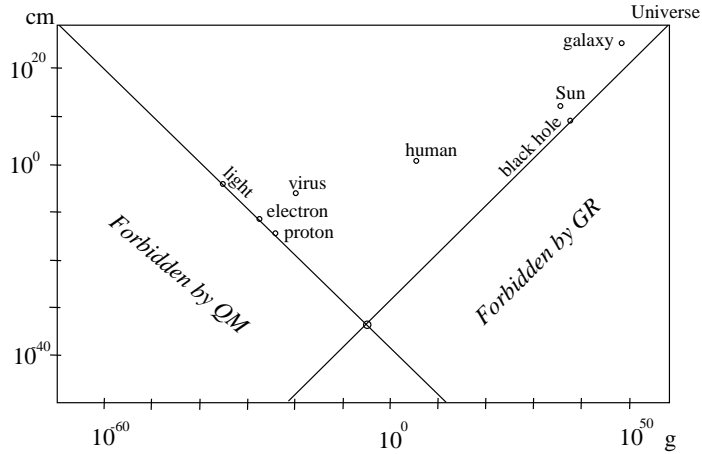}\]
\caption{Range of physical phenomena, which lie in the wedge
region with us in the middle. Log plots are mass-energy v size}
\end{figure}

Note that as theoretical physicists we are not {\em obliged} to
explain why the above should be a crude model for some of the
structure of physics. We need only observe that it is a non-empty
postulate with predictive and explanatory power. However, it is
possible to speculate a little further and come to some
philosophical conclusions. I will do this now. Why the principle
of self-duality? Why such a central role for Fourier theory? The
answer I believe is that something very general like this
underlies the very nature of what it means to do science. The
basic problem was proposed by Plato and comes under the heading of
`Plato's cave'. Namely, how can one tell the difference between
reality and its representation or shadow. The modern answer
according to \cite{Ma:pri} was that one should consider not one
representation (or one angle of projection in Plato's cave) but
{\em all} of them. Their collection is as much a valid
mathematical structure as the original, as explained in above.
Thus suppose that some theorist puts forward a theory in which
there is an actual group $G$ say `in reality' (this is where
physics differs from mathematics) and some experimentalists
construct tests of the theory and in so doing they routinely build
representations or elements of $\hat G$. They will end up
regarding $\hat G$ as `real' and $G$ as merely an encoding of
$\hat G$. The two points of view are in harmony because
mathematically (in the appropriate context)
\eqn{GGhat}{ G\isom\hat{\hat{G}}.}
So far so good, but through the interaction and confusion between
the experimental and theoretical points of view one will
eventually have to consider both, i.e. $G\times
\hat G$ as real. But then the theorists will come along and say
that they don't like direct products, everything should interact
with everything else, and will seek to unify $G,\hat G$ into some
more complicated irreducible structure $G_1$, say. Then the
experimentalists build $\hat G_1$ ... and so on. This is a kind of
engine for the evolution of Science. For example, if one regarded,
following Newton, that space $\R^n$ is real, its representations
$\hat \R^n$ are derived quantities ${\bf p}=m\dot {\bf x}$. But
after making diverse such representations one eventually regards
both $\bf x$ and $\bf p$ as equally valid, equivalent via Fourier
theory. But then we seek to unify them and introduce the CCR
algebra (\ref{heis}). And so on. Note that this is not intended to
be a historical account but a theory for how things could have
gone in an ideal case without the twists and turns of human
ignorance.

This is the plausibility reason that something {\em like} the
principle of representation-theoretic self-duality should be
observed. We have given arguments above that there is at least a
correlation between the mathematical structure of self-dual
structures and the progressive theories of physics from their
birth in `logic' to the projected forthcoming complete theory of
everything. It should at least provide a guide to the properties
that should be central in unknown theories of everything, such as
what have become fashionable to call `M-theory'. Now what if this
kind of self-duality of structure was not only observed (in a
crude form) but something like it, perhaps along with some other
key postulates, actually fully characterised the structure of the
allowed theories of physics? This is not out of the question given
what we have said above about the level of generality already
reached. It would be like giving a list of things that we expect
from a complete theory -- such as renormalisability,
CPT-invariance, etc., except that we are considering such general
versions of these `constraints' that they are practically what it
means to be a group of people following the scientific method. If
this really pins down the ultimate theory then it would mean that:
\begin{itemize}
\item The ultimate
theory of physics may be no more and no less than a self-discovery
of the constraints in thinking that are taken on when one decides
to look at the world as a physicist.
\end{itemize}

This is not at all the usual view of Nature as blindly `out there'
and is what I meant by a new philosophical foundation for
theoretical physics. As big ideas go it is basically Kantian or
Hegelian as opposed to the more conventional reductionist one that
most physicists take for granted. The difference is that whereas
Kant could only speculate, science backed by experiment, may
actually be coming to the same conclusion in the not impossibly
far future. It is important to note that this would not mean that
physics is arbitrary or random any more than the different
possible manifolds `out there' are arbitrary. The space of all
possible manifolds up to equivalence has a deep and rich structure
and feels every bit as real to anyone who studies it; but it is a
mathematical reality `created' when we accept the axioms of a
manifold. So what we are saying is that there is not such a
fundamental difference between mathematical reality and physical
reality. The main difference is that mathematicians are aware of
the axioms while physicists tend to discover them `backwards' by
theorising from experience. I call this subjunctive point of view
{\em relative realism}\cite{Ma:pri}. In it, we experience reality
through choices that we have forgotten about at any given moment.
If we become aware of the choice the reality it creates is
dissolved or `unconstructed'. On the other hand, the reader will
say that the possibility of the theory of manifolds -- that the
game of manifold-hunting could have been played in the first place
-- is itself a reality, not arbitrary. It is, but at a higher
level: it is a concrete fact in a more general theory of possible
axiom systems of this type. To give another example, the reality
of chess is created once we chose to play the game. If we are
aware that it is a game, that reality is dissolved, but the rules
of chess remain a reality although not within chess but within the
space of possible board games. This gives a tree-like or
hierarchical structure of reality. Reality is experienced as we
look down the tree while `awareness' or enlightenment is achieved
as we look up the tree. When we are born we take on millions and
millions of assumptions or rules through communication, which
creates our day to day perception of reality, we then spend large
parts of our lives questioning and attempting to unconstruct these
assumptions as we seek understanding of the world. And from this
perspective the fact that life appears somewhere near the middle
of Figure~10, apart from the obvious explanation that phenomena
become simpler as we approach the boundaries hence most complex in
the middle so this is statistically where life would develop, has
a different explanation: we created our picture of physical
reality around ourselves and so not surprisingly we are near the
middle. We may in effect have painted ourselves in a box by taking
on certain assumptions about how to go about looking at the world.
This would not be a bad thing but rather a statement about the
origin of the laws of physics. I do doubt that it is ever going to
be as simple as all that, but it is something to think about on a
rainy day in the next millennium.

\subsection*{Acknowledgments}
The author is a Reader and Royal Society University
Research Fellow at QMW and a Senior Research Fellow during 1999-2001 at
the Department of Applied Maths and Theoretical Physics, University of
Cambridge, England, where some of the work was completed.

\baselineskip 15pt

\end{document}